\documentclass[12pt]{elsarticle}

\usepackage[T1]{fontenc}
\usepackage{color}
\usepackage{array}
\usepackage{colortbl}
\usepackage{nomencl}
\makenomenclature
\usepackage{amssymb}
\usepackage{amsmath}
\usepackage{etoolbox}
\usepackage{mathtools}
\usepackage{epsfig}
\usepackage{tabularx}
\usepackage{multirow}
\usepackage{subfigure}
\usepackage{babel}
\usepackage{psfrag}
\usepackage{paralist}
\usepackage{setspace}
\usepackage{comment}
\usepackage{morefloats}
\usepackage{vale}
\usepackage{enumitem}
\usepackage[normalem]{ulem}
\usepackage[hidelinks,pdfencoding=auto, psdextra]{hyperref}
\hypersetup{pdfauthor={Name}}
\usepackage{bm}
\usepackage[a4paper, margin=1in]{geometry}
\PassOptionsToPackage{pdftex}{graphicx}
\usepackage{tikz}
\usepackage{algorithmicx,algpseudocode}
\usepackage[compact]{titlesec}
\usepackage{titlesec}
\usepackage{url}
\usepackage{amsthm}

\newcounter{bla}

\journal{International Journal of Heat and Mass Transfer}

\begin{document}

\begin{frontmatter}

\title{The Wulff bio--heat transfer model revisited: directional blood enthalpy transport, the biological Peclet number, and implications for laser-induced thermal therapy}
\author[ancona]{Valerio D'Alessandro \corref{cor}}
\ead{v.dalessandro@univpm.it}
\cortext[cor]{Corresponding author}
\author[ancona]{Matteo Falone }
\ead{m.falone@staff.univpm.it}
\author[ancona]{Luca Giammichele}
\ead{l.giammichele@staff.univpm.it}
\author[ancona]{Renato Ricci}
\ead{ricci@univpm.it}

\address[ancona]{Dipartimento di Ingegneria Industriale e Scienze Matematiche,\\ Universit\`a Politecnica delle Marche, Via Brecce Bianche 12, 60131 Ancona (AN), Italy}

\begin{abstract}
Bio--heat transfer models play a fundamental role in predicting temperature fields during laser--induced thermal therapy (LITT).
Among continuum bio--heat transfer models, the Pennes equation remains the most widely adopted formulation. However, by representing blood perfusion as an isotropic volumetric source, it neglects the directional transport of thermal energy associated with blood flow. The Wulff model overcomes this limitation by incorporating blood averaged enthalpy transport directly into the heat flux. Despite its physical significance, the derivation of the Wulff formulation and the assumptions required to obtain its governing equation remain only briefly discussed in the original work.\\
%
The aim of the present work is twofold. First, the physical formulation originally proposed by Wulff is revisited in order to clarify its derivation and the assumptions required to obtain the governing equation. In particular, the physical origin of the modified heat flux is investigated and an equivalent formulation with an independent metabolic source term is discussed. Second, the resulting model is assessed against the classical Pennes formulation through representative laser--induced thermal therapy benchmark problem.
The numerical results show that accounting for blood flow directionality may substantially alter the predicted temperature field and the extent of thermal damage. A dimensionless analysis further leads to the definition of a biological Peclet number, which quantifies the magnitude of directional blood enthalpy transport relative to thermal diffusion.
Finally, a simple extension accounting for venous stasis is proposed, mitigating the excessive temperature rise predicted by the original Wulff formulation.
%
\end{abstract}

\begin{keyword}
Wulff model; Pennes model;  biological Peclet number; Laser-induced thermal therapy.
\end{keyword}

\end{frontmatter}

\section{Introduction} \label{sec:intro}
Hyperthermia based ablatation techniques for tumor treatment  raise the local temperature
beyond normal physiological levels. Actually, these methods are designed 
to cause irreversible cellular damage, triggering apoptosis and coagulative 
necrosis in tumor tissues.
%
%
Laser--induced thermal therapy (LITT) represents one of the most promising hyperthermia based ablative techniques owing to its minimally invasive nature and its ability to deliver highly localized thermal energy. Accurate thermal modelling is therefore essential for predicting temperature distributions and the resulting tissue damage, providing a valuable tool for treatment planning and optimization.
For instance, Saccomandi et al.~\cite{Saccomandi2012}
proposed a theoretical model to evaluate LITT outcomes in pancreatic tumors, while Khosravirad and Ayani~\cite{KHOSRAVIRAD2023106837} 
analyzed the thermal impact of laser treatment on breast cancer. Additionally, Andreozzi et al.~\cite{Andreozzi2024} demonstrated the potential of laser ablation in inducing necrosis in prostate cancer.\\ 
%
%
%
%
Numerous bio--heat transfer formulations have been proposed for the simulation of LITT. Despite this variety, most continuum models still rely on the assumption of isotropic blood perfusion.
In this context, the Pennes bio--heat equation,~\cite{Pennes:1948}, remains the most widely adopted continuum model for describing heat transfer in biological tissues. However, it neglects the directional transport of thermal energy associated with blood flow.
To account for this effect, Wulff~\cite{Wulff:1974} proposed the first continuum bio--heat model incorporating the directional transport of thermal energy associated with the blood flow. Although the model has been cited for decades, the original paper provides only a concise derivation of the governing equation, so that some of the underlying physical assumptions remain only partially clarified.
%
For this reason in the present work, Wulff model formulation is revisited to examine the physical consistency of its metabolic source term, an aspect that has received little attention in the bio--heat transfer literature, including recent review articles,~\cite{YUAN2025107499}.
Therefore, one of the main aims of this paper is to critically assess the predictive capabilities of the Pennes and Wulff bio--heat models through a rigorous comparison on benchmark problems available in the literature.
Moreover, the theoretical formulation of the Wulff bio--heat model is revisited, clarifying the physical origin of the directional blood enthalpy transport term. The introduction of an independent metabolic heat source is also discussed. \\  
On this basis, the numerical implementation of the Wulff formulation is first validated against an analytical solution for a one-dimensional steady-state problem. Subsequently, both models are employed to simulate goldnanoparticles (GNPs) assissted laser-induced thermal therapy of an axisymmetric skin melanoma,~\cite{Ren2017,Falone2024}, allowing a direct comparison of their predictions. 
The obtained results reveal that blood flow directionality can significantly influence both the predicted temperature field and the resulting extent of thermal damage. This behavior is rationalized through the proposed biological Peclet number, which quantifies the effect of directional blood enthalpy transport.\\
Finally, since existing Wulff formulations do not account for the progressive reduction of blood flow caused by thermal damage, a modified formulation including venous stasis is proposed, a phenomenon that, to the best of the authors' knowledge, has previously been incorporated only into Pennes bio--heat equation.\\
The paper is organized as follows Sec.~\ref{sec:gov} treats the description of governing equations. Sec.~\ref{sec:num} briefly describes the numerical approximation techniques. Sec.~\ref{sec:res} is devoted to the results and Sec.~\ref{sec:concl} contains the conclusions.

\section{Governing equations}\label{sec:gov}
\subsection{Bio--heat transfer models}
\noindent Energy conservation equation for heat conduction in a biological tissue can be written as: 
\begin{equation}\label{eq:cond_gen}     
        \rho_t c_{p,t} \frac{\partial T_t}{\partial t} = -\NABLA \cdot {\bf q}_t + G~,
\end{equation}
where $\rho_t$ and $c_{p,t}$ denote the tissue density and specific heat capacity, respectively; $T_t$ represents the tissue temperature, ${\bf q}_t$ is the heat flux vector, and $G$ denotes the internal heat generation term.\\
The Pennes bio--heat model employs Fourier’s law to describe the heat flux vector. To account for the effect of blood perfusion, a specific source/sink term is introduced through the quantity $G_b$, defined as $G_b = \rho_{b} \omega_b c_{p,b} \left( T_{a,0} - T_t \right)$,~\cite{Pennes:1948}. 
Furthermore, metabolic heat generation is
accounted for through the volumetric heat source terms $G_m$.
Therefore, the Pennes bio--heat equation reads
\begin{equation}\label{eq:cond_pennes2}
\rho_t c_{p,t}\frac{\partial T_t}{\partial t}
=
\NABLA\cdot\left(\lambda_t\NABLA T_t\right)
+
\rho_b\omega_b c_{p,b}(T_{a,0}-T_t)
+
G_m~.
\end{equation}
In eq.~\ref{eq:cond_pennes2}, quantities identified by the subscript $b$
refer to blood properties, where $\omega_b$ is the blood perfusion rate,
$c_{p,b}$ is the blood specific heat capacity, and $T_{a,0}$ is the arterial
blood temperature.\\
It is worth emphasizing that biological tissues are mainly composed of water, which
may undergo vaporization owing to the temperature increase induced by laser heating.
This phase change significantly affects the thermophysical properties introduced above.
To account for these temperature dependent variations, the enthalpy method proposed by
Abraham and Sparrow,~\cite{ABRAHAM20072537}, is adopted. Therefore, the volumetric
heat capacity is expressed as:
\begin{equation}\label{eq:enthalpy}
    \rho_t c_{p,t}=\left\{
    \begin{array}{@{}rl}
      \left( \rho_l c_{p,l} \right)_t & \text{if }~~0~\mathrm{^\circ} C <T \leq 99~\mathrm{^\circ} C\\
\frac{h_{fg}C_{w,t}}{\Delta T}& \text{if }~~99~\mathrm{^\circ} C <T \leq 100~\mathrm{^\circ} C\\
      \left( \rho_g c_{p,g} \right)_t & \text{if }~~ T > 100~\mathrm{^\circ} C~.
    \end{array} \right.
\end{equation}
In eq.~\ref{eq:enthalpy}, $\rho_l$ and $c_{p,l}$ denote the density and specific
heat capacity of the tissue in the liquid phase, whereas $\rho_g$ and $c_{p,g}$
represent the corresponding properties of the vaporized tissue. Furthermore,
$h_{fg}$ is the product of the latent heat of vaporization of water and the water
density at $100~^\circ\mathrm{C}$, $C_{w,t}$ is the tissue water content, and
$\Delta T$ denotes the temperature interval over which the phase change takes
place, equal to $1~^\circ\mathrm{C}$.\\
A second formulation for blood perfusion considered in the present work is the model proposed
by Wulff,~\cite{Wulff:1974}. Unlike the Pennes equation, Wulff's formulation
incorporates the directional transport of thermal energy by blood directly into the
heat flux vector. The original paper introduces the modified heat flux expression without an explicit derivation. For completeness, its physical basis is revisited here by considering the energy balance for the blood phase, thereby making explicit the assumptions underlying the subsequent derivation.
In the Wullf's model the total heat flux is expressed as the sum of the conductive heat
flux through the tissue and the enthalpy flux associated with blood motion:
\begin{equation}\label{eq:wulff_heat_flux}
{\bf q}_t
=
-\lambda_t\NABLA T_t
+
\rho_b{\bf v}_h h_b~.
\end{equation}
Here, ${\bf v}_h$ denotes the Darcy velocity of blood, while the specific blood
enthalpy is written as

\begin{equation} \label{eq:wulff_enthalpy}
h_b=
\int_{T_0}^{T_b}
c_{p,b}(T_b')\,dT_b'
+
\frac{P}{\rho_b}
+
\Delta H_f(1-\epsilon)~.
\end{equation}
in which $T_0$ and $T_b$ represent the reference temperature for the specific enthalpy evaluation and the blood temperature respectively, $P$ denotes the pressure, $\Delta H_f$ is the specific enthalpy of metabolic reactions and $\epsilon$ represents the metabolic efficiency.\\
The additional term in eq.~\ref{eq:wulff_heat_flux} is specifically introduced to account the enthalpy transport related to the blood flow, therefore the heat flux vector may be written as:
\begin{equation}
\label{eq:qtot}
{\bf q}_t={\bf q}_f+{\bf q}_{b\rightarrow t}~,
\end{equation}
where ${\bf q}_f$ denotes the conductive heat flux within the tissue and
${\bf q}_{b\rightarrow t}$ represents the net energy transported by the
blood.
The latter contribution can be obtained by considering the integral energy
balance over a representative control volume $\Omega_b$ occupied by the blood,
where $\partial \Omega$ is its boundary partially shared with the surrounding tissue.
Although the thermal problem addressed in this work is unsteady, the
characteristic time associated with blood convection through the
microvascular network is several orders of magnitude smaller than the
characteristic thermal diffusion time in the surrounding tissue.
Consequently, the blood phase may be handled as locally in a quasi--steady
state. Under this assumption, the unsteady term appearing in the blood energy equation
can be neglected. Furthermore, owing to the relatively low blood velocity,
both kinetic energy variations and viscous dissipation are negligible with
respect to the enthalpy transport. The integral energy balance therefore
reduces to
\begin{equation}
\label{eq:blood_balance}
\oint_{\partial\Omega_b}
\left(
\rho_b {\bf v}_h h_b
+
{\bf q}_{t\rightarrow b}
\right)\cdot{\bf n}\,dS
=
0~,
\end{equation}
where ${\bf q}_{t\rightarrow b}$ denotes the heat flux exchanged between the
surrounding tissue and the blood. Eq.~\ref{eq:blood_balance} states that the enthalpy transported by the blood is balanced by
the heat exchanged through the vessel wall. 
Therefore, it is possible to write the following equation:
\begin{equation}
\label{eq:qblood}
{\bf q}_{b\rightarrow t}
= -{\bf q}_{t\rightarrow b}
= \rho_b {\bf v}_h h_b~,
\end{equation}
which represents the averaged enthalpy flux associated with the blood motion.
Substituting eq.~\ref{eq:qblood} into eq.~\ref{eq:qtot} yields the modified
heat flux expression proposed by Wulff.
It should be emphasized that the second term in
eq.~\ref{eq:wulff_heat_flux} is not a constitutive law analogous to Fourier's
law. It represents a continuum approximation of the net enthalpy
transport produced by the microcirculation after spatial averaging over the
vascular network. Consequently, the validity of the model relies on the
assumptions underlying the averaging procedure and on the existence of a
meaningful local averaged blood velocity field.\\
Substituting eq.~\ref{eq:wulff_heat_flux} into the local energy conservation
equation and invoking blood mass conservation,
$\NABLA\cdot( \rho_b {\bf v}_h)=0$, together with the assumption of negligible
mechanical work, yields
\begin{equation}\label{eq:wulff_2}
	\rho_t c_{p,t} \frac{\partial T_t}{\partial t} = \NABLA \cdot \left(  \lambda_t \NABLA T_t \right) - \rho_b {\bf v}_h \cdot \NABLA \left[ 
											\int_{T_0}^{T_b} c_{p,b} \left(T_b'\right) dT_b' + \Delta H_f \left( 1- \epsilon \right) \right]
\end{equation}
which can be further simplified by assuming temperature independent  thermophysical properties and a uniform $T_0$:
\begin{equation}\label{eq:wulff_3}
\rho_t c_{p,t} \frac{\partial T_t}{\partial t} = \NABLA \cdot \left(  \lambda_t \NABLA T_t \right) - \rho_b {\bf v}_h \cdot \left( 
		c_{p,b} \NABLA T_b - \Delta H_f \NABLA \epsilon
	\right)~.
\end{equation}
At this stage, to complete the derivation two additional closure assumptions are required. First, local thermal
equilibrium between blood and tissue is assumed,~\emph{i.e.} $T_b=T_t$, consistently with
the hypothesis that blood reaches thermal equilibrium with the surrounding tissue
within the microcirculation. Second, the term
$\rho_b\Delta H_f{\bf v}_h\cdot \NABLA\epsilon$ is identified with the volumetric
metabolic heat generation rate, $G_m$, yielding:
\begin{equation}\label{eq:wulff_4}
\rho_t c_{p,t} \frac{\partial T_t}{\partial t} = \NABLA \cdot \left(  \lambda_t \NABLA T_t \right) - \rho_b c_{p,b} {\bf v}_h \cdot  
		 \NABLA T_t  + G_m~.
\end{equation}
While the first assumption is commonly adopted in continuum models for bio--heat transfer, the second one
deserves particular attention. Indeed, the identification of
$\rho_b\Delta H_f{\bf v}_h\cdot\NABLA \epsilon$ with a prescribed volumetric heat
source does not directly follow from the conservation equations and may lead to
non-physical spatial variations of the metabolic efficiency along the blood-flow
direction. 
To illustrate this point, consider a one-dimensional blood flow, ${\bf v}_h=v_h{\bf i}$,
for which the above expression becomes
\begin{equation}
\label{eq:qm_wulff}
G_m=\rho_b v_h \Delta H_f \frac{\partial\epsilon}{\partial x}~,
\end{equation}
assuming ${\bf v}_h\cdot\NABLA\epsilon>0$. If the local metabolic heat generation rate is approximately uniform along the capillary bed, eq.~\ref{eq:qm_wulff} yields
\begin{equation}
\epsilon(x)=\epsilon_0+\frac{G_m}{\rho_b v_h \Delta H_f}x~,
\end{equation}
which predicts a monotonic increase of the metabolic efficiency along the blood flow direction. This trend is opposite to the expected physiological behavior, since blood progressively loses oxygen and nutrients while flowing through the microcirculation, leading to a reduction rather than an increase in the capability of sustaining metabolic reactions. Therefore, the orginal  Wulff formulation may produce paradoxical predictions unless the blood velocity field and the spatial variation of the metabolic efficiency are prescribed consistently.\\
Actually, from the physical point of view metabolic heat generation and blood convection are different phenomena. Metabolic heat is produced locally by cellular activity, while blood serves only to thermal energy transport throughout the tissue.
Therefore, the metabolic and convective contributions can be introduced independently into the energy balance. In particular, metabolic heat generation may be treated as a prescribed volumetric source, as originally proposed by Pennes, while the blood enthalpy is restricted to its sensible contribution. Under the assumption of local thermal equilibrium this alternative formulation leads to exactly the same governing equation as Wulff model. The only difference lies in the physical interpretation: by removing the metabolic contribution from the blood enthalpy in eq.~\ref{eq:wulff_enthalpy}, the governing equation is recovered without introducing the metabolic efficiency gradient, thereby avoiding the conceptual inconsistencies discussed above.\\
It is interesting to note as Wulff model solves the principal physical inconsistency of the Pennes model by introducing directional enthalpy transport. However, this improvement comes at the price of requiring a local averaged blood velocity field, whose determination is generally more difficult than prescribing a perfusion rate.\\
Lastly, it is highly important to stress as under the hypotesis of uniform thermal conductivity Pennes' model
equation, eq.~\ref{eq:cond_pennes2}, can be re--written in its dimensionless
form as follows:
\begin{equation}\label{eq:pennes_adim}
\frac{1}{\mathrm{Fo}}\frac{\partial \widehat{T_t}}{\partial \widehat{t}} =
\widehat{\nabla}^2 \widehat{T_t}  - \left(\tilde{m}_p L \right)^2   \widehat{T_t} + \widehat{G}_m~,
\end{equation}
where $\widehat{T_t} = \left(T_t - T_{a,_0}\right)/\left(T_i - T_{a,_0}\right)$ is the dimensionless temperature, defined with respect to the tissue's initial temperature $T_i$. By contrast, $\mathrm{Fo}$ denotes the Fourier number, $\widehat{t}$ the dimensionless time, $L$ represents the characteristic length of the problem, and $\widehat{G}_m$ is the dimensionless counterpart of $G_m$ as defined in eq.~\ref{eq:cond_pennes2}. 
Finally, $\tilde{m}_p$ represents a model parameter defined in this case as:

\begin{equation}\label{eq:m-pennes}
 \tilde{m}_p = \sqrt{ \frac{\rho_b c_{p,b} \omega_b}{\lambda_t} }~.
\end{equation}
Similarly, Wulff's model equation, eq.~\ref{eq:wulff_4}, can be re--arranged
in a dimensionless fashion as follows:
\begin{equation}\label{eq:wulff_adim}
\frac{1}{\mathrm{Fo}}\frac{\partial \widehat{T_t}}{\partial \widehat{t}} =
\widehat{\nabla}^2 \widehat{T_t}  - \left(\tilde{m}_w L \right)^2  {\bf \widehat{v}_h} \cdot \widehat{\NABLA} \widehat{T_t} + \widehat{G}_m 
\end{equation}
where $\widehat{G}_m$ is the dimensionless expression of ${G}_m$, while ${\bf \widehat{v}_h}$ is the blood Darcy velocity unit vector and $\tilde{m}_w$ is defined as below:
\begin{equation}\label{eq:m-wulff}
 \tilde{m}_w = \frac{\rho_b c_{p,b} \left| {\bf {v}_h}  \right|  }{\lambda_t}~.
\end{equation}
\subsection{Laser--heating and thermal damage}
The laser-induced heating effect, if present, is included by means of a specific source terms $G_L$ in both Pennes and Wulff models.
The presence of gold nanoparticles (GNPs) makes the tumor an absorption-dominated
medium, as discussed in Soni et al.~\cite{SONI201470}. Under these conditions, light attenuation
can be accurately described by the Beer--Lambert law,~\cite{Jacques1993}:
\begin{equation}
\label{eq:cond1_1}
I(s)=I_0\exp(-\mu s)~,
\end{equation}
where $I_0$ is the incident laser intensity and $\mu = \mu_a + \mu_s$ is the effective attenuation
coefficient. The assorbtion and scattering coefficients are defined as
\begin{equation}
\label{eq:coeff}
\begin{aligned}
\mu_a &= \mu_{a,t}+\mu_{a,G}~,\\
\mu_s &= \mu_{s,t}+\mu_{s,G}~.
\end{aligned}
\end{equation}
Here, $\mu_{a,t}$ and $\mu_{s,t}$ denoting the absorption and scattering coefficients
of the tissue, respectively, whereas $\mu_{a,G}$ and $\mu_{s,G}$ account for the corresponding
contributions of the GNPs. Following~ Dombrovsky et al.~\cite{DOMBROVSKY20115459}, these quantities
are given by
\begin{equation}
\label{eq:coeff1}
\begin{aligned}
\mu_{a,G} &=0.75\,f_v\,\frac{Q_a}{r}~,\\
\mu_{s,G} &=0.75\,f_v\,\frac{Q'_s}{r}~,
\end{aligned}
\end{equation}
where $f_v$ is the nanoparticle volume fraction, $r$ is the nanoparticle radius,
$Q_a$ is the absorption efficiency factor and $Q'_s$ is the transport scattering
efficiency factor. Note that in eq.~\ref{eq:cond1_1} the variable $s$ denotes the laser propagation direction which, in this work coincides with the negative $y$--axis, see Fig.~\ref{fig:scheme}. Thus, the volumetric heat generation due to laser
absorption is given by
\begin{equation}
G_L=\mu I(s)~.
\end{equation}
Thermal damage is quantified by the Arrhenius model,~\cite{Henriques:1947}:
\begin{equation}
\label{eq:damage}
\Omega(t)
=
A
\int_0^t
\exp\!\left(
-\frac{E_a}{RT(t')}
\right)~dt'~,
\end{equation}
where $A$ is the frequency factor, $E_a$ is the activation energy and $R$ is the
universal gas constant. Tissue is assumed to undergo irreversible thermal damage
when $\Omega>1$,~\cite{Henriques:1947}.\\
In the context of Pennes model, the reduction of blood perfusion caused by vascular stasis is accounted for through
the model proposed by He \textit{et al.}~\cite{HE2004}. Defining the degree of
vascular stasis as
\[
\beta_s=1-\exp(-\Omega)~,
\]
the local blood perfusion rate becomes
\begin{equation}
\label{eq:omega}
\omega_b
=
\omega_{b,0}(1-\beta_s)~,
\end{equation}
where $\omega_{b,0}$ refers to the physiological perfusion rate.
On the other hand, it is worth noting that in the original Wulff formulation, the convective transport associated with blood flow remains unaffected by the progressive vascular damage induced during thermal therapy. However, venous stasis is expected to reduce the effective blood perfusion and, consequently, the energy convective transport. 
To account for this effect, the present work proposes scaling the convective term by the attenuation factor $\exp(-\Omega)$, consistently with the reduction in blood perfusion adopted for the Pennes model. The resulting governing equation reads:
\begin{equation}
\label{eq:wulff_stasis}
\rho_t c_{p_t}\frac{\partial T_t}{\partial t}
=
\NABLA\cdot\left(\lambda_t\NABLA T_t\right)
-
\left(\rho_b c_{p,b}\mathbf{v}_{h} \cdot \NABLA T_t \right)\exp\left({-\Omega}\right)
+
G_m~,
\end{equation}
where the proposed attenuation should be interpreted as a phenomenological correction accounting for the progressive reduction of directional blood enthalpy transport caused by vascular stasis.

%
%
%
%
%

\section{Numerical approximation}\label{sec:num}
The governing equations were solved using a specifically developed numerical solver based on the \texttt{foam-extend} v.5.0 library, a fork of the well-known OpenFOAM code, which is strongly focused on integrating community contributions. The solution technique is based on a second-order accurate, colocated, cell-centered finite volume method. All diffusive terms were approximated with second-order accuracy, while a second-order implicit Euler method (BDF-2) was used for time integration. 
The solver is capable of handling two distinct computational domains, which are subsequently solved in a conjugate fashion using a fully coupled solution technique.
This approach is used for cases involving healthy tissue and a tumor portion, where each part is considered a medium with uniform thermophysical properties. Linear systems are solved using a stabilized bi-conjugate gradient method, preconditioned with the Cholesky method and they are considered converged when the residuals reached machine precision.

\section{Results}\label{sec:res}
\subsection{Validation}\label{sec:val}
\begin{figure}[htbp]
 \centering
 {\includegraphics[width=0.45\textwidth, trim=2mm 2mm 2mm 2mm, clip]{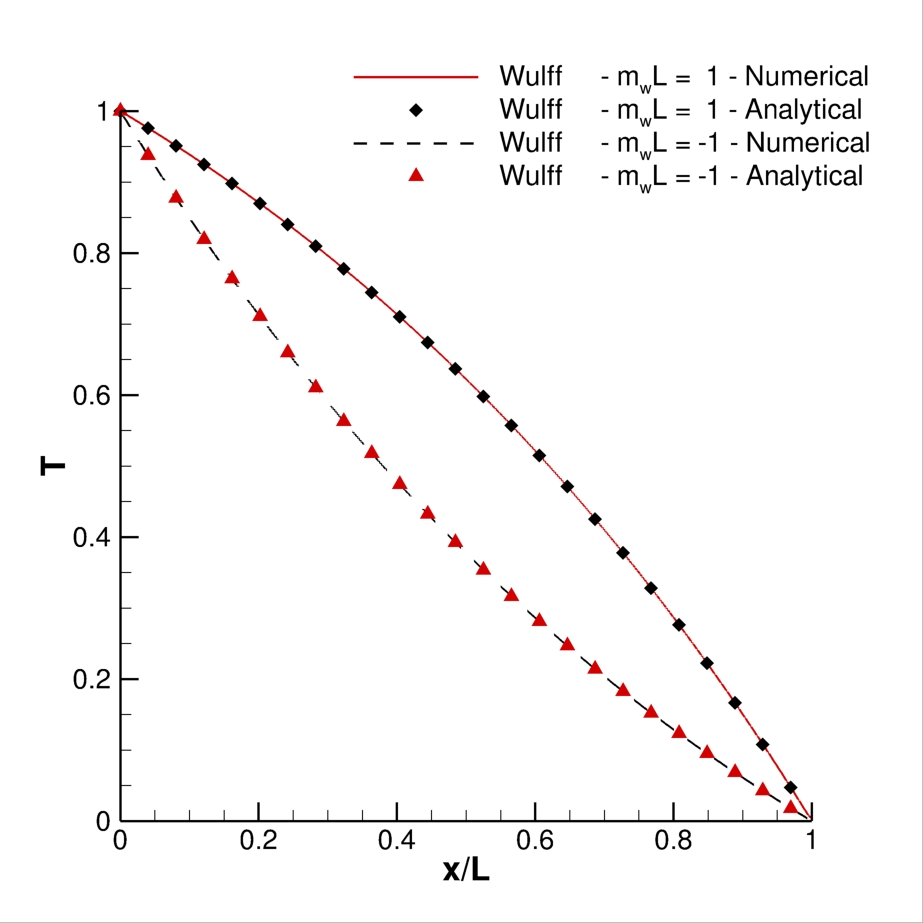}}
\caption{Dimensionless temperature distribution along the slab thickness.}
\label{fig:validation}
\end{figure}

The numerical solver developed for the Wulff model is firstly benchmarked against an analytical solution derived for steady-state heat transfer in a one-dimensional slab. The corresponding solver for the Pennes model was previously validated in our former work,~\cite{Falone2024}.\\
Wulff model equation, eq.~\ref{eq:wulff_4}, under steady-state conditions without metabolic contributions reads:
\begin{equation}\label{eq:wulff-an1}
\frac{d^2 T_t}{d x^2} - \tilde{m}_w\frac{d T_t}{d x}  = 0~.
\end{equation}
The boundary conditions applied are: $T\left(x = 0\right) = T_c$ and $T\left(x = L_s\right) = T_s$, where $L_s$ denotes the slab thickness. Consequently, the solution of eq.~\ref{eq:wulff-an1}, obtained via the method of separation of variables, is given by:
\begin{equation}
    T_t(x) \;=\; T_c \;-\; (T_c - T_s)\,
    \frac{\exp  \left( {\tilde{m}_w x} \right)- 1}{\exp \left( {\tilde{m}_w L}\right) - 1}~
\end{equation}
where the temperature can be also expressed in dimensionless form as: 
$\widehat{T} = \left(T_t - T_{s}\right)/\left(T_c - T_{s}\right)$~.\\
As shown in Fig.~\ref{fig:validation}, our numerical solver exhibits excellent agreement with the analytical solution and for this reason it can be considered sufficiently reliable for replicating cases reported in the literature. Specifically, $\tilde{m}_w L$ was set equal to $\pm 1$ for the considered configuration in order to asses also the solver reliability with respect of blow flow directionality.\\
\subsection{Axisymmetric melanoma surrounded by healthy tissue}
\begin{figure}[htbp]
 \centering
 \subfigure[\label{fig:geom} Computational domain.]
 {\includegraphics[width=0.45\textwidth]{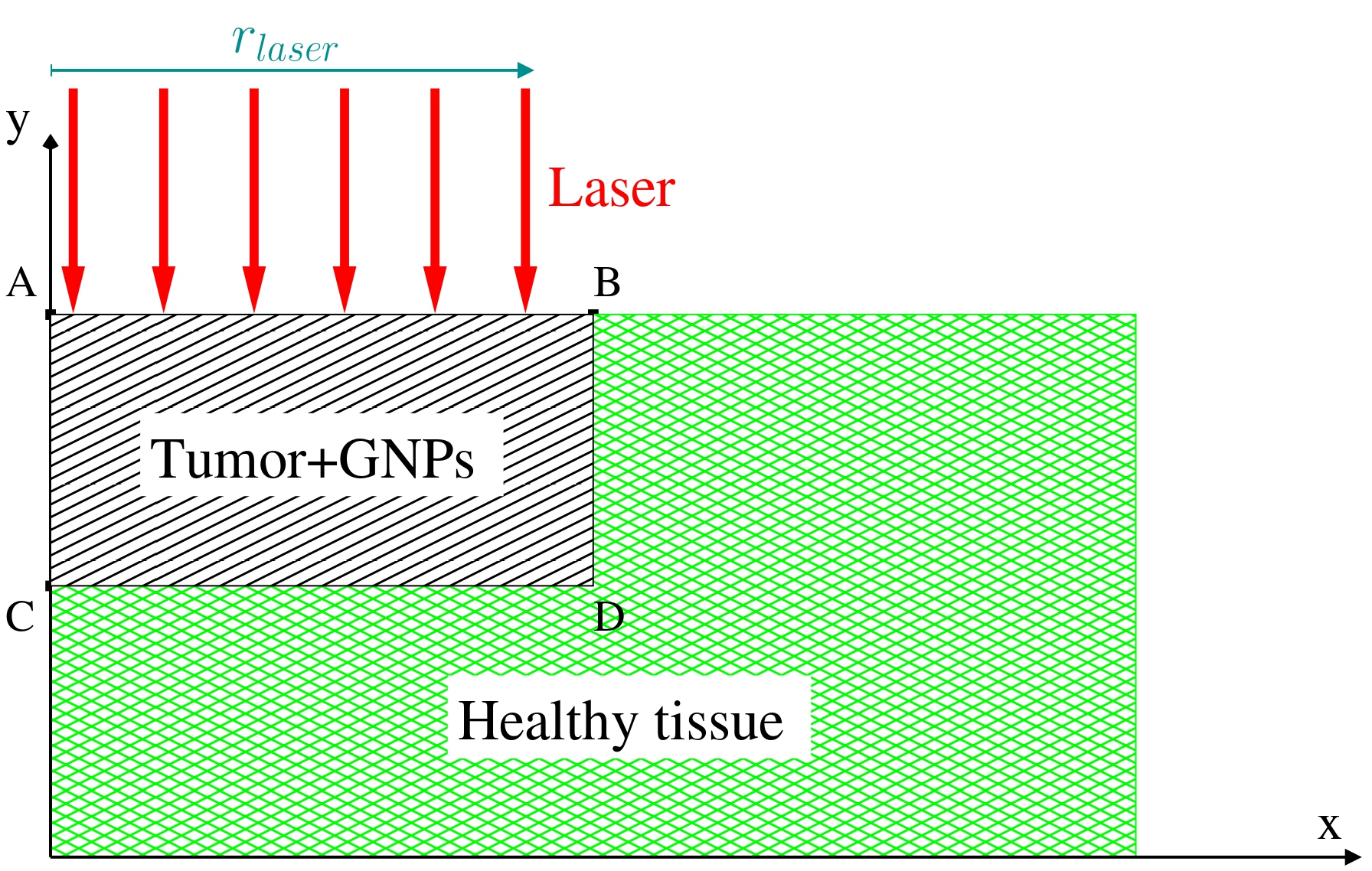}}
\hfill
  \subfigure[\label{fig:laser} Laser administration strategy.]
 {\includegraphics[width=0.45\textwidth,trim=2mm 2mm 2mm 2mm, clip]{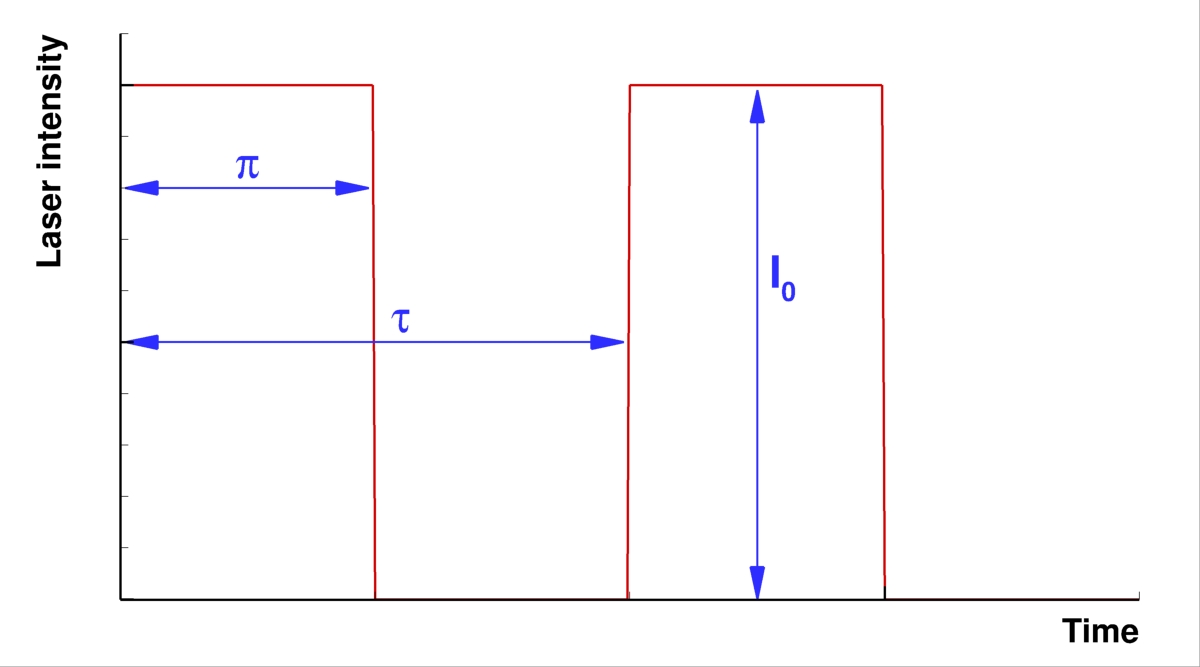}}
\caption{Schematic of the investigated configuration.}
\label{fig:scheme}
\end{figure}

\begin{figure}[htbp]
 \centering
 \subfigure[\label{fig:Ta1000-mp_mw} Point A.]
 {\includegraphics[width=0.45\textwidth, trim=2mm 2mm 2mm 2mm, clip]{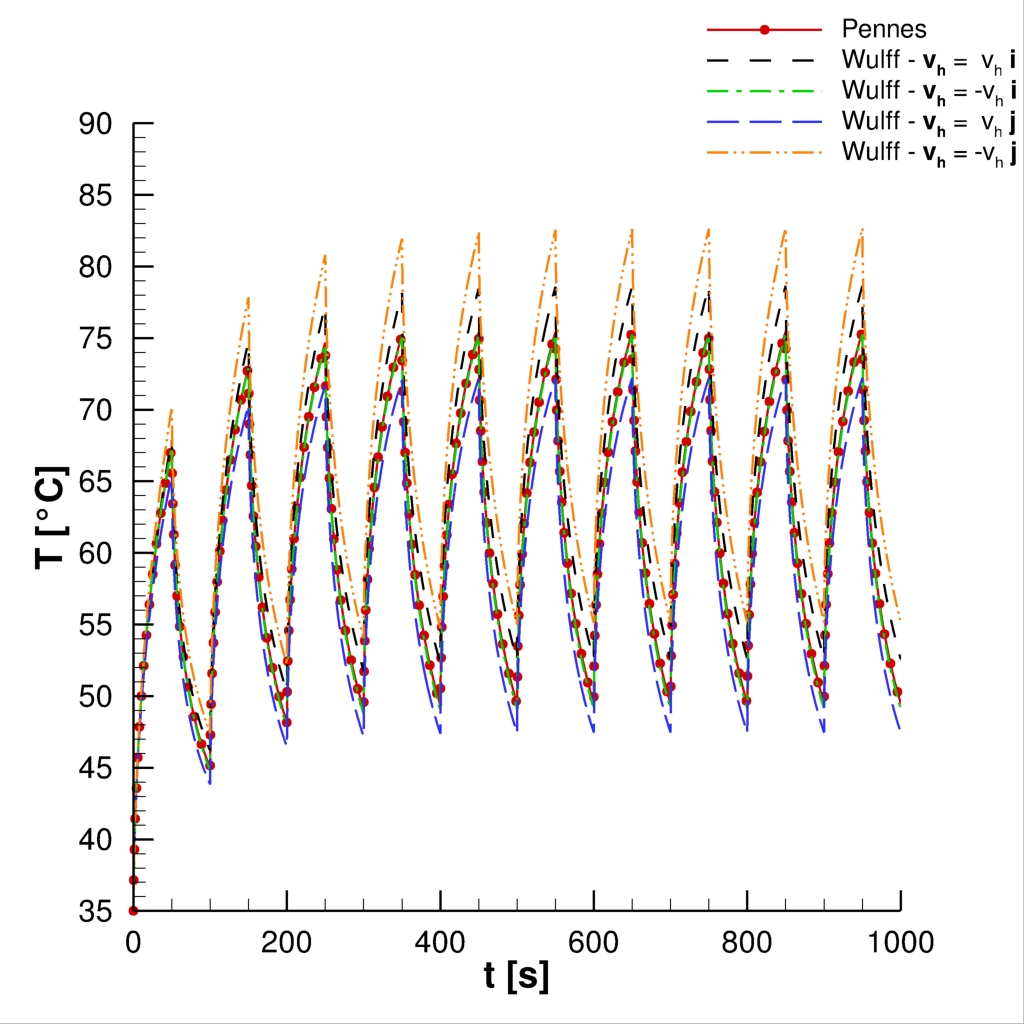}}
\hfill
  \subfigure[\label{fig:Tb1000} Point B.]
 {\includegraphics[width=0.45\textwidth, trim=2mm 2mm 2mm 2mm, clip]{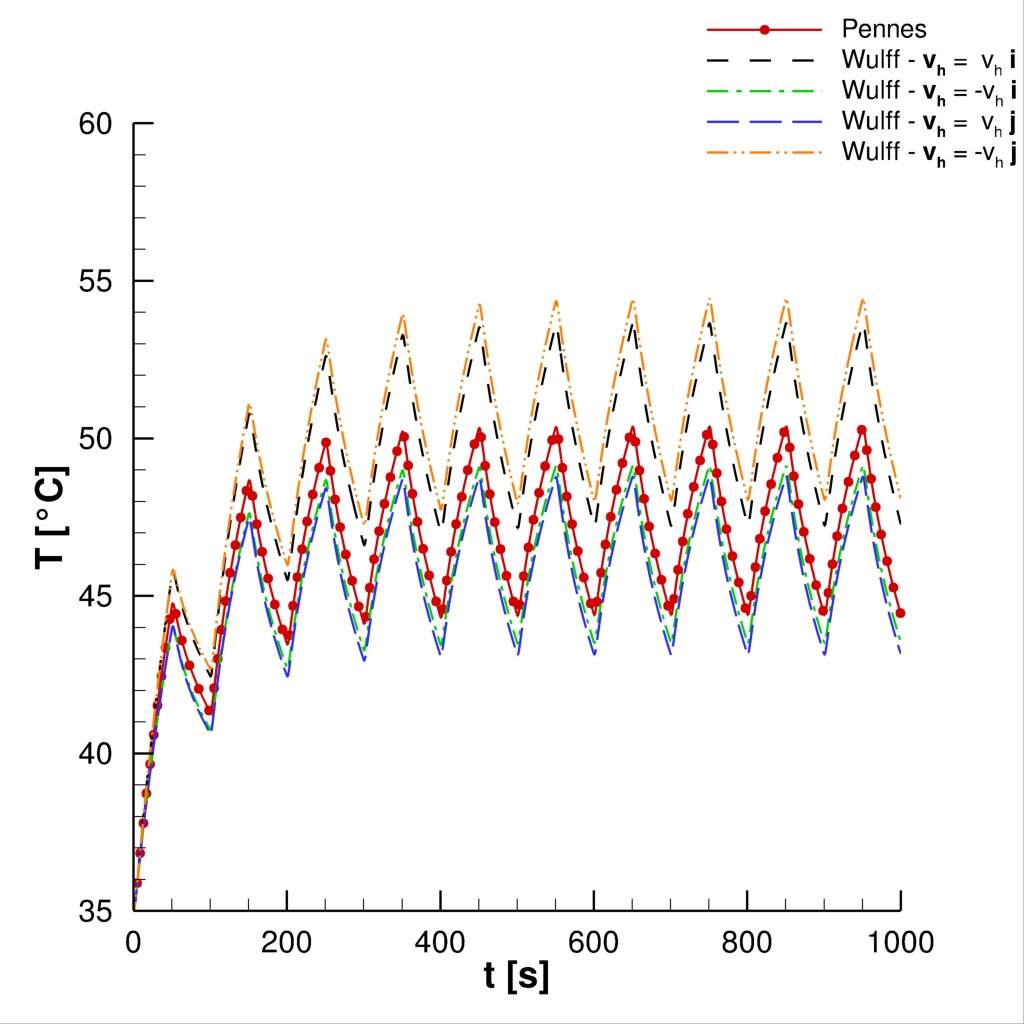}}
\caption{Time-history of temperature in point A and B, $\tilde{m}_w  = \tilde{m}_p$.}
\label{fig:Ta_m}
\end{figure}

\begin{figure}[htbp]
 \centering
 \subfigure[\label{fig:Ta1000} Point C.]
 {\includegraphics[width=0.45\textwidth, trim=2mm 2mm 2mm 2mm, clip]{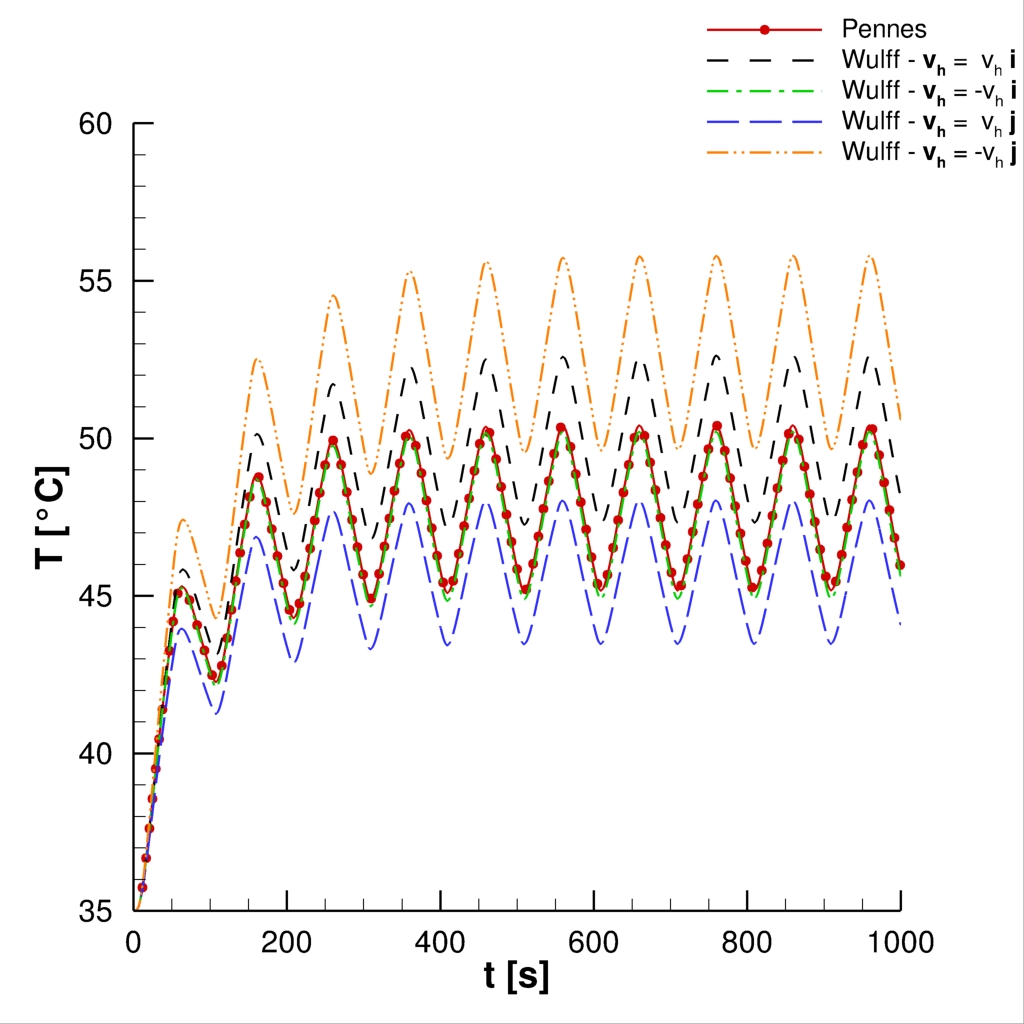}}
\hfill
  \subfigure[\label{fig:Td1000-mp_mw} Point D.]
 {\includegraphics[width=0.45\textwidth, trim=2mm 2mm 2mm 2mm, clip]{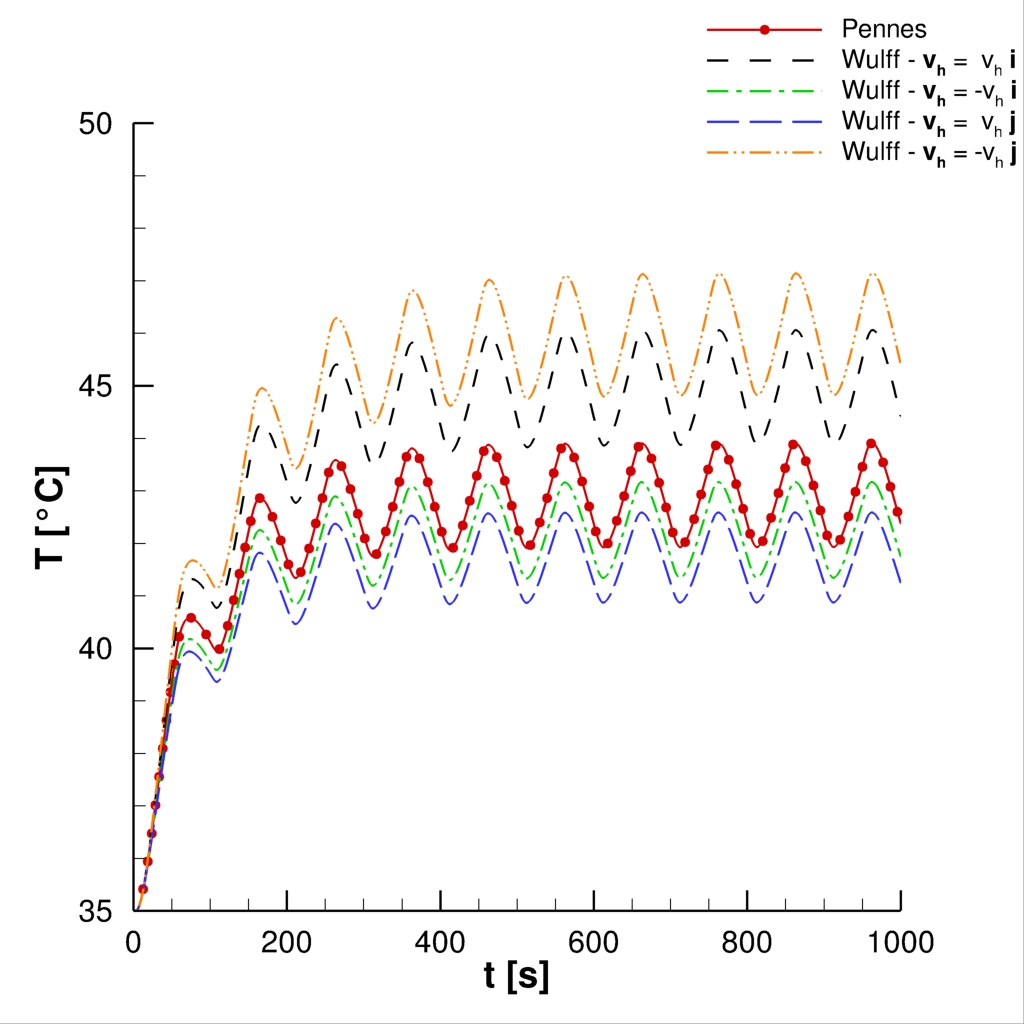}}
\caption{Time-history of temperature in point C and D, $\tilde{m}_w  = \tilde{m}_p$.}
\label{fig:Ta_m}
\end{figure}

\begin{figure}[htbp]
 \centering
  \subfigure[\label{fig:fPennes} Pennes.]
 {\includegraphics[width=0.45\textwidth, trim=2mm 2mm 2mm 2mm, clip]{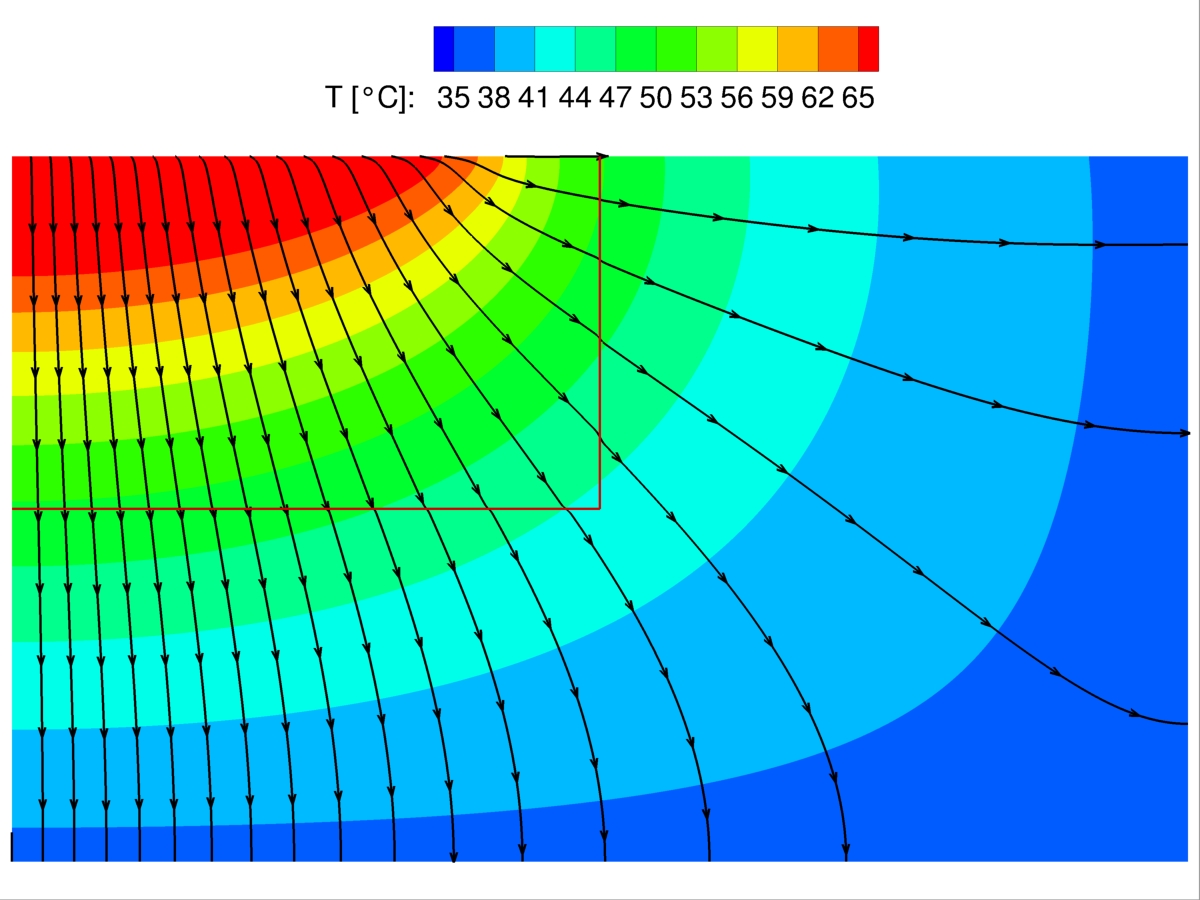}}
\hspace{5cm}
 \subfigure[\label{fig:fUpos} Wulff - $\mathbf{v}_h=v_h~\mathbf{\hat{i}}$~.]
 {\includegraphics[width=0.45\textwidth, trim=2mm 2mm 2mm 2mm, clip]{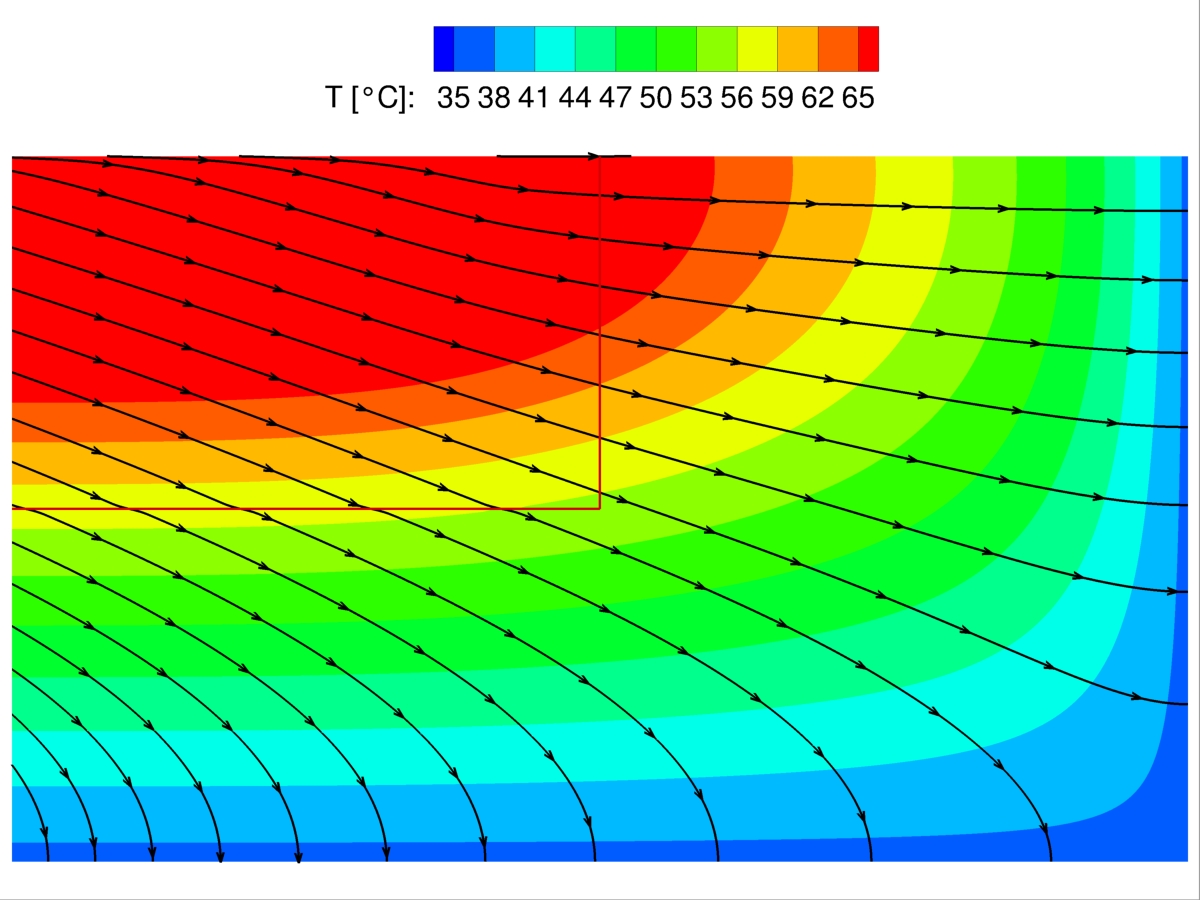}}
\hfill
  \subfigure[\label{fig:fUneg} Wulff - $\mathbf{v}_h=-v_h~\mathbf{\hat{i}}$~.]
 {\includegraphics[width=0.45\textwidth, trim=2mm 2mm 2mm 2mm, clip]{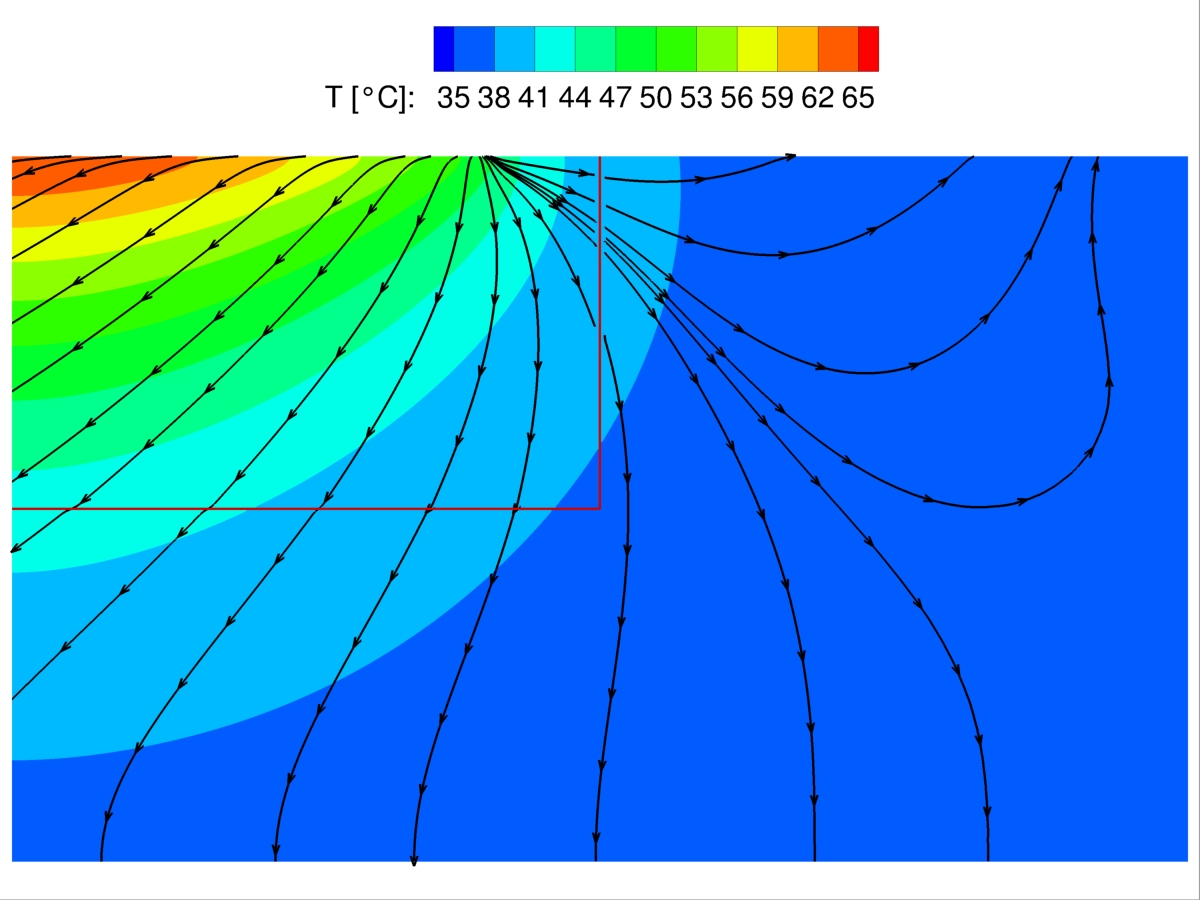}}
\medskip
 \subfigure[\label{fig:fvpos} Wulff - $\mathbf{v}_h=v_h~\mathbf{\hat{j}}$~.]
 {\includegraphics[width=0.45\textwidth, trim=2mm 2mm 2mm 2mm, clip]{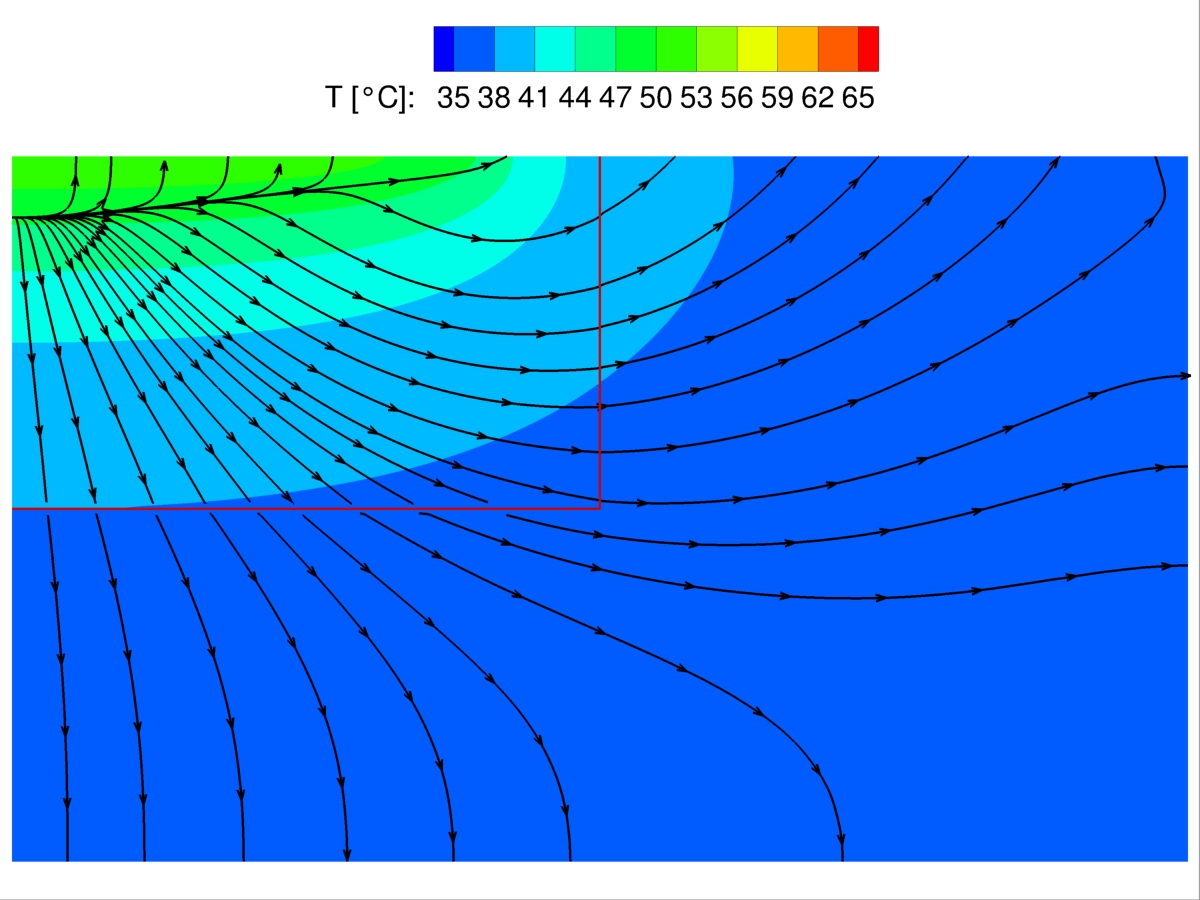}}
\hfill
  \subfigure[\label{fig:fVneg} Wulff - $\mathbf{v}_h=-v_h~\mathbf{\hat{j}}$~.]
 {\includegraphics[width=0.45\textwidth, trim=2mm 2mm 2mm 2mm, clip]{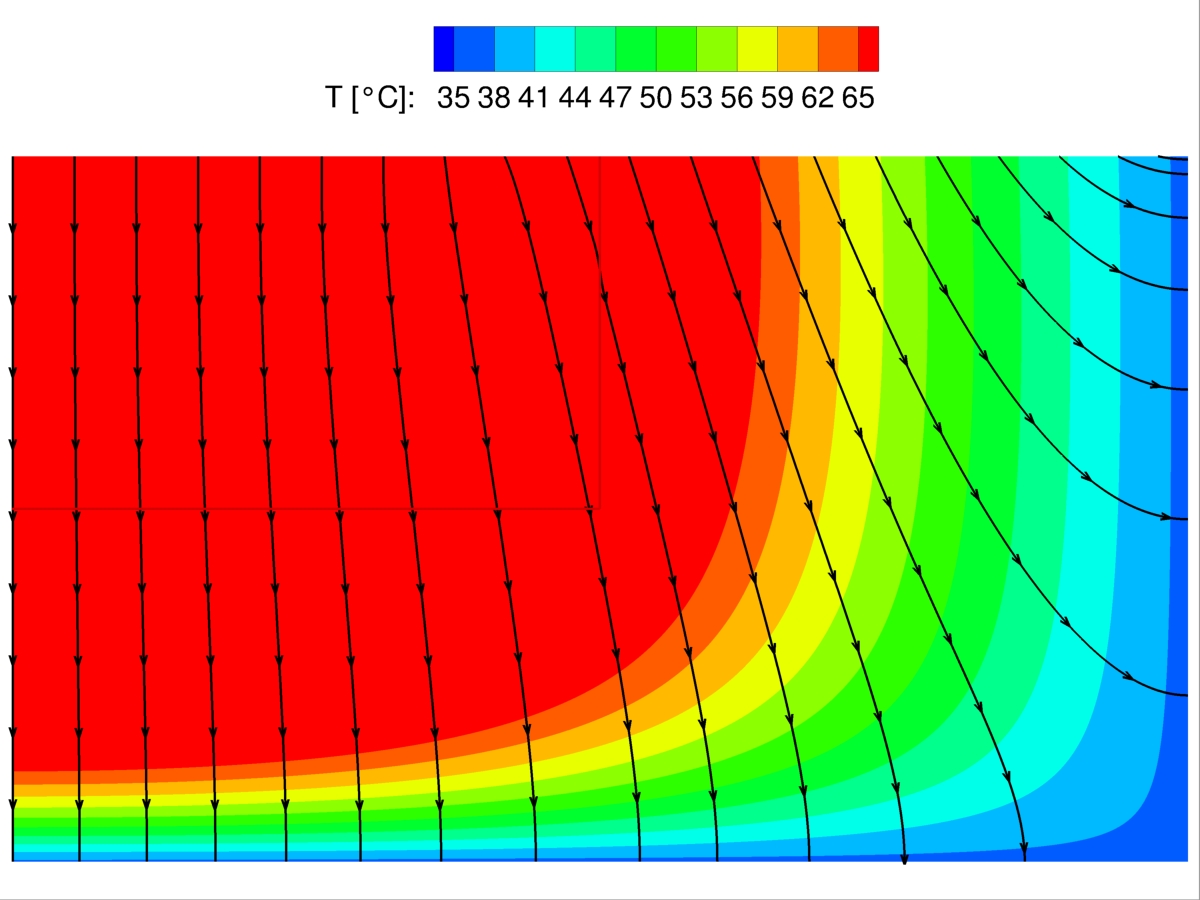}}
\caption{Temperature contour fields and heatflux lines - $t=950~s$.}
\label{fig:flux}
\end{figure}

\begin{figure}[htbp]
 \centering
 \subfigure[\label{fig:Ta1000} Point A.]
 {\includegraphics[width=0.45\textwidth, trim=2mm 2mm 2mm 2mm, clip]{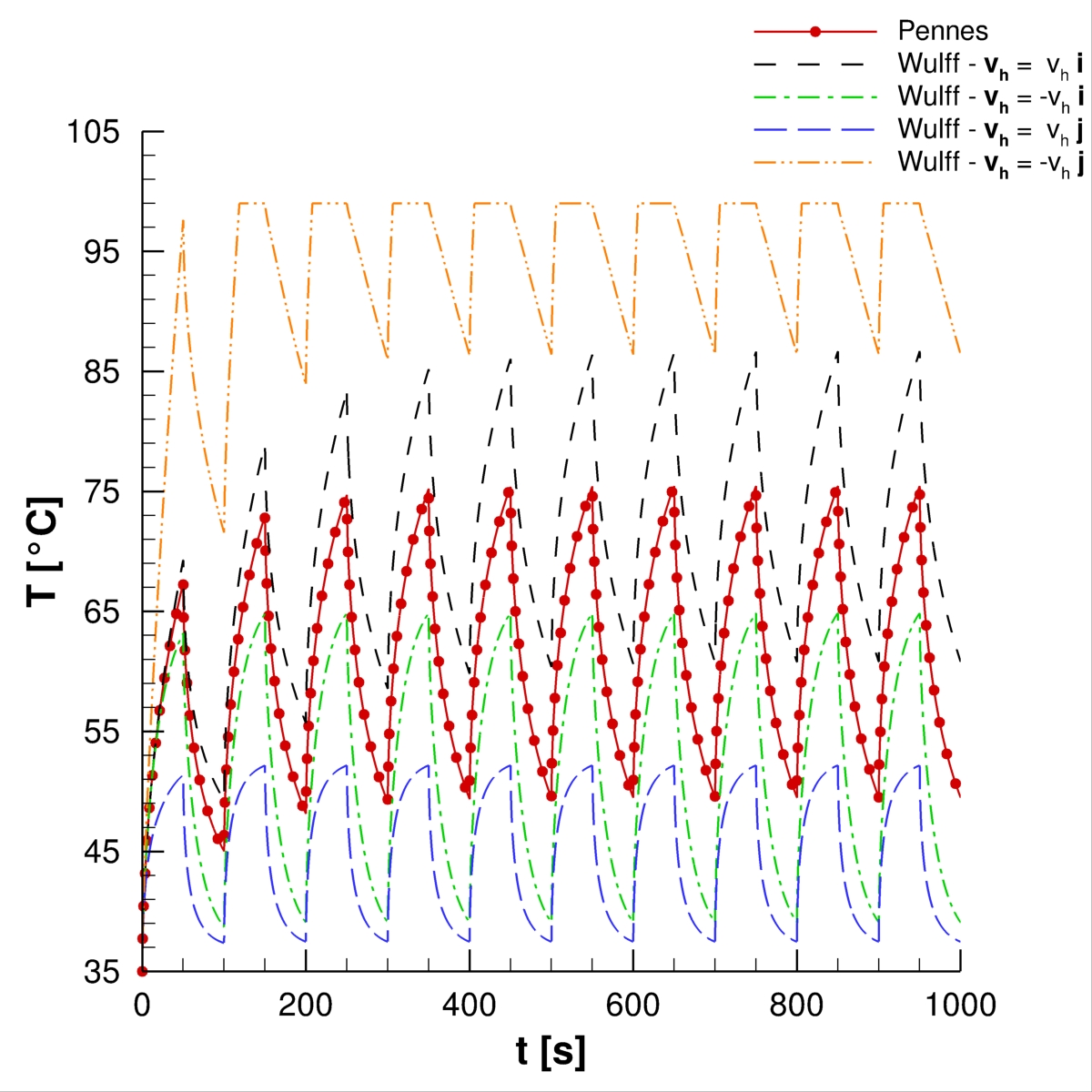}}
\hfill
  \subfigure[\label{fig:Tb1000} Point B.]
 {\includegraphics[width=0.45\textwidth, trim=2mm 2mm 2mm 2mm, clip]{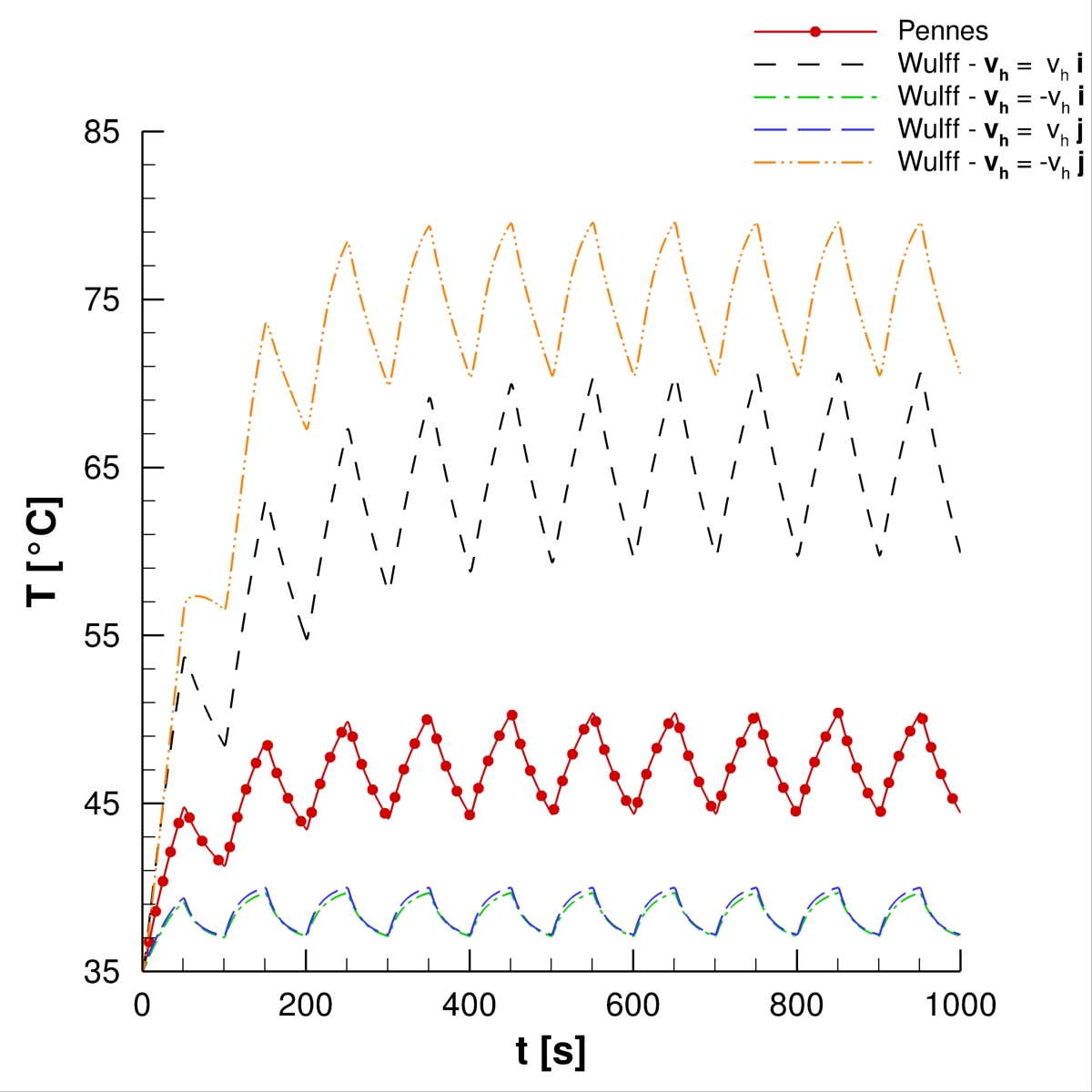}}
\caption{Time-history of temperature in point A and B, $\tilde{m}_w  = 10~\tilde{m}_p$.}
\label{fig:Tab_m}
\end{figure}

\begin{figure}[htbp]
 \centering
 \subfigure[\label{fig:Tc1000} Point C.]
 {\includegraphics[width=0.45\textwidth, trim=2mm 2mm 2mm 2mm, clip]{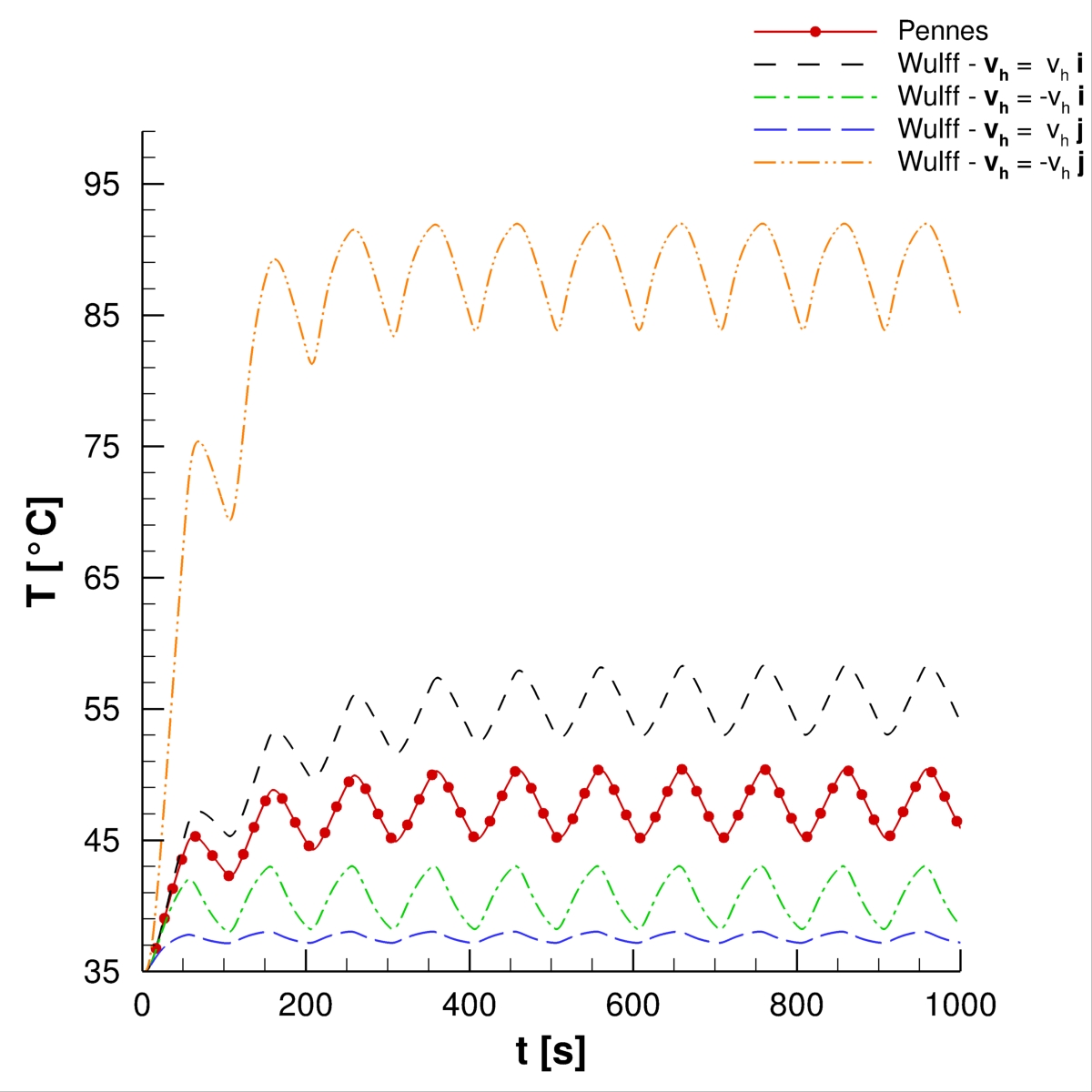}}
\hfill
  \subfigure[\label{fig:Td1000} Point D.]
 {\includegraphics[width=0.45\textwidth, trim=2mm 2mm 2mm 2mm, clip]{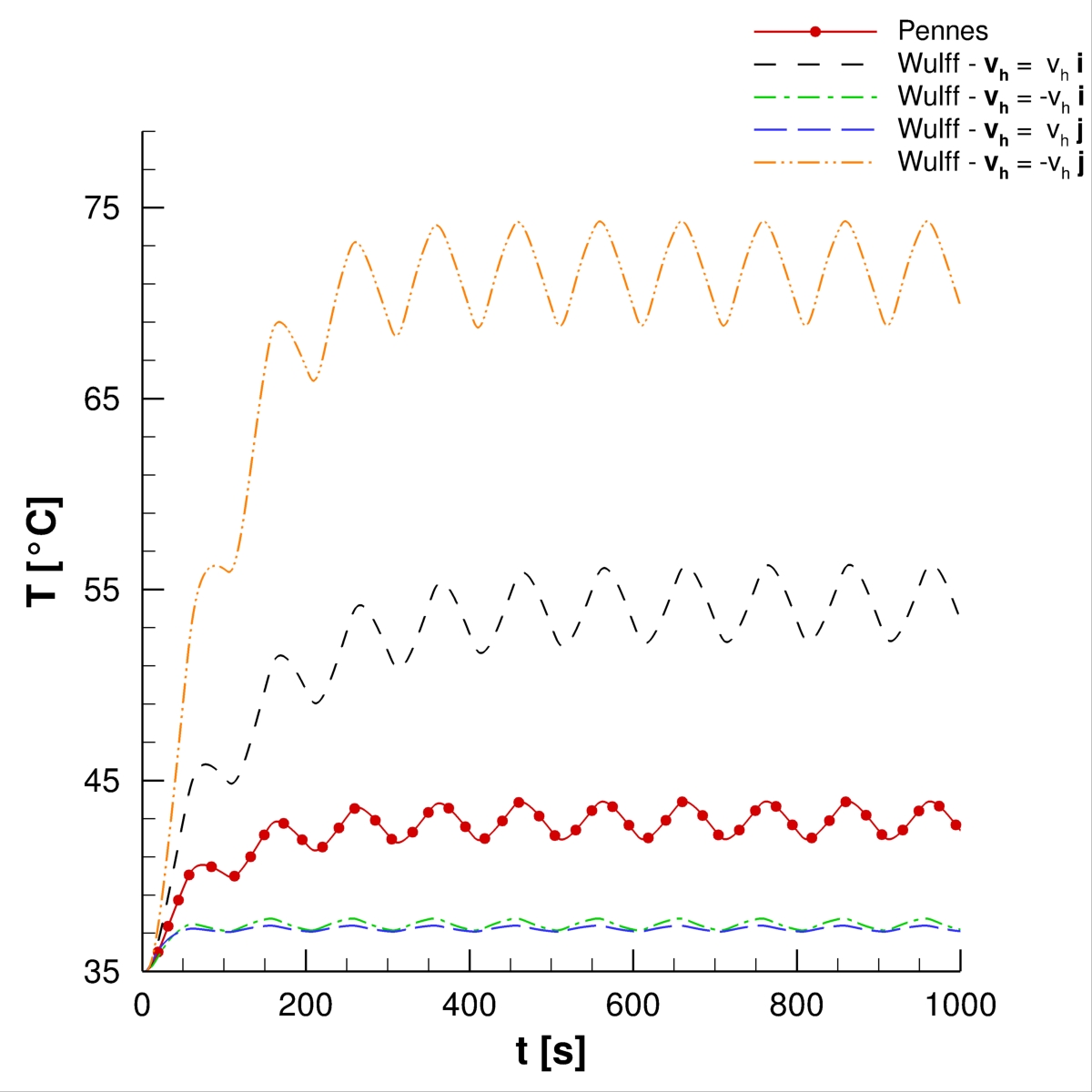}}
\caption{Time-history of temperature in point C and D, $\tilde{m}_w  =  10~\tilde{m}_p$}
\label{fig:Tcd_m}
\end{figure}

The effect of blood flow directionality with respect to the heat flux is then evaluated, considering its impact on LITT for skin tumors in terms of standard parameters related to heat transfer problems~\emph{i.e.}:
\begin{inparaenum}[(i)]
  \item temperature time--history in significant points;
  \item temperature contour fields;
  \item heat flux lines.
\end{inparaenum}
Moreover, additional parameters were investigated to quantify both the effectiveness of the treatment and the extent of undesired effects. In particular, we refer to the treatment efficiency, $\varepsilon_{L}$, defined as the percentage of tumor volume in which $\Omega \ge 1$. 
Conversely, $U_L$ represents the percentage of healthy tissue that undergoes irreversible damage.\\
The computations reported in the following were conducted on the well--known configuration proposed by Ren et al.,~\cite{Ren2017}. 
It consists of an axisymmetric problem in which a melanoma is surrounded by healthy tissue, as schematized in Fig.~\ref{fig:geom}.
In particular, the tumoral zone extension is considered to be 5 mm in x direction and 3 mm in y direction, while the adjacent volume extends up to x = 10 mm and y = 6 mm. 
The tumor is assumed to be infused with gold nanorods characterized by an effective radius of 11.43 nm and an aspect ratio of 3.9. The nanoparticles are considered to be uniformly distributed within the tumor tissue, with a volume fraction $f_v$ equal to $10^{-5}$.
Thermal ablation of cancer cells is achieved through exposure to a laser beam with a radius of $r_{\text{laser}} = 4.2$ mm, operating at a wavelength of 808 nm (Gallium Arsenide laser).\\
The laser induced heating is modeled using Beer’s law, in accordance with our previous studies~\cite{Falone2024, Falone_2025}.
Specifically, a pulsed irradiation protocol is adopted, Fig.~\ref{fig:laser}.  The nominal intensity $I_0$ equal to $\mathrm{0.75~W/cm^2}$ is provided to the tumor for a pulse width $\pi=50$ s followed by a rest time that completes the basic cycle having a period $\tau$ of 100 s. The treatment has a duration of 1000 s.\\
The thermophysical properties of healthy tissue are assumed to be
$\rho_t = 10^3~\mathrm{kg/m^3}$,
$c_{p,t} = 4.2 \cdot 10^3~\mathrm{J/kg\,K}$,
and
$\lambda_t = 0.5~\mathrm{W/m\,K}$.
For the tumoral tissue, the same specific heat is adopted, whereas the density and thermal conductivity are taken as $ 1.1 \cdot 10^3~\mathrm{kg/m^3}$ and
$0.55~\mathrm{W/m\,K}$, respectively.
Blood properties are
$\rho_b = 10^3~\mathrm{kg/m^3}$
and
$c_{p,b} = 4.2 \cdot 10^3~\mathrm{J/kg\,K}$.
A uniform volumetric metabolic heat generation rate of
$G_m = 1091~\mathrm{W/m^3}$
is employed.\\
Finally, the optical properties are taken from the literature~\cite{Ren2017,Falone2024}. In particular, the healthy tissue absorption and scattering coefficients are
$\mu_{a,t}=2~\mathrm{m^{-1}}$
and
$\mu_{s,t}=650~\mathrm{m^{-1}}$,
whereas for the tumor loaded with gold nanorods they are
$\mu_a=12100~\mathrm{m^{-1}}$
and
$\mu_s=50~\mathrm{m^{-1}}$.\\
For all the computations described below, a fully structured mesh with $9.6\cdot 10^4$ uniformly spaced elements having a characteristic cell's length, $L_c$, of $\mathrm{ 2.5 \cdot 10^{-4}~m}$, was used to discretize the computational domain and a time-step sizes $\Delta t$ equal to $\mathrm{10^{-4}~s}$ was employed basing on the authors' previous work,~\cite{Falone2024}. 
A body temperature of $T_B = 37~\mathrm{^\circ C}$ was applied to the right and bottom boundaries. A convective heat transfer boundary condition, characterized by a convective heat transfer coefficient of $h = 5~\mathrm{W/m^2K}$ and an external temperature of $T_\infty = 25~\mathrm{^\circ C}$, was imposed on the upper boundary.
Finally, a symmetry condition was applied to the left boundary, and an initial temperature of $35~\mathrm{^\circ C}$ was set throughout the entire computational domain.\\
%
%
In this paper, a campaign of simulations was carried out by considering one Pennes configuration and two different Wulff configurations. For the Pennes model, $\widetilde{m}_p$ was set equal to 91.562 according to the thermophysical properties proposed by Ren et al.~\cite{Ren2017}. For the Wulff model, two values of $\widetilde{m}_w$ were considered. In the first configuration, $\widetilde{m}_w=\widetilde{m}_p=91.562$, allowing a direct comparison between the two formulations under identical dimensionless perfusion conditions. In the second configuration, the Darcy velocity was set to $v_h=10^{-4}$ m/s, obtained by assuming a blood velocity of $5\cdot10^{-4}$ m/s,~\cite{Stucker:1996,Fagrell:1977}, and a tissue porosity of 0.2,~\cite{Liu2024}. This leads to $\widetilde{m}_w=915.62$, corresponding to a physiologically realistic Darcy velocity.
Four different orientations were considered for $\mathbf{v}_h$, on the basis of the possible capillaries direction in human skin,~\cite{Stucker:1996,Fagrell:1977}:
\begin{inparaenum}[(i)]
  \item horizontal, positive direction  $\mathbf{v}_h=v_h~\mathbf{\hat{i}}$; 
  \item horizontal, negative direction  $\mathbf{v}_h=-v_h~\mathbf{\hat{i}}$; 
  \item vertical, positive direction  $\mathbf{v}_h=v_h~\mathbf{\hat{j}}$; 
  \item vertical, negative direction  $\mathbf{v}_h=-v_h~\mathbf{\hat{j}}$. 
\end{inparaenum}
\\
Fig.~\ref{fig:Ta1000-mp_mw}--\ref{fig:Td1000-mp_mw} show the temporal evolution of temperature at monitoring points A--D (see Fig.~\ref{fig:scheme}) obtained with the Pennes and Wulff formulations for the case $\widetilde{m}_w=\widetilde{m}_p$.
Even under identical values of the dimensionless perfusion parameter, the two models predict markedly different thermal responses.
When the velocity is aligned with the laser propagation direction (${\bf v}_h=-v_h{\bf j}$), higher temperatures are obtained throughout the treatment. By contrast, reversing the flow direction (${\bf v}_h=v_h{\bf j}$) leads to a marked reduction in temperature. For transverse blood flow, two distinct thermal responses emerge. Notably, the case $\mathbf{v}_h=v_h\mathbf{i}$ produces higher temperatures, although they remain lower than those predicted for the configuration aligned with the laser beam. The opposite flow direction ($\mathbf{v}_h=-v_h\mathbf{i}$) leads to temperature fields that are generally more similar to those predicted by the Pennes formulation. Overall, these results are not surprising, since the Wulff formulation explicitly accounts for the direction of blood flow through the advective term.\\
Tab.~\ref{tab:E-m1} summarizes the treatment efficiency and the extent of collateral thermal damage for the case $\tilde{m}_w$=$\tilde{m}_p$. Tab.~\ref{tab:E-m1} highlights that also the treatment efficiency predicted by the Wulff formulation is markedly influenced by the orientation of the blood flow. Depending on the direction of the velocity, the predicted efficiency varies from $50.2\%$ to $90.7\%$, whereas the Pennes model provides a single value of $63.8\%$. In contrast, the extent of healty tissue thermal damage remains negligible for all the investigated configurations. These results indicate that, even when the same dimensionless perfusion parameter is imposed, the directional contribution introduced by the Wulff formulation has a noticeable impact on the predicted treatment efficiency, while the extent of healty tissue thermal damage remains limited.
\\
\begin{table}
\caption{Treatment efficiency and undesired effect, $\widetilde{m}_w=\widetilde{m}_p=91.562$}
\label{tab:E-m1}
\centering\begin{tabular}{ccccccc}
\hline\noalign{\medskip}
& Pennes & Wulff & Wulff & Wulff & Wulff\\
&& $\mathbf{v}_h=v_h~\mathbf{\hat{i}}$ & 
$\mathbf{v_h}=-v_h~\mathbf{\hat{i}}$ &
$\mathbf{v_h}=v_h~\mathbf{\hat{j}}$ &
$\mathbf{v_h}=-v_h~\mathbf{\hat{j}}$\\
\hline\noalign{\medskip}
$\mathrm{\varepsilon_L}$~~[\%] & 63.8 & 82.9 & 59.8 & 50.2 & 90.7\\\hline\noalign{\medskip}
$\mathrm{U_{L}}$~[\%]    &   0.0 &  0.2 &  0.0 & 0.0  &  2.8\\\hline\noalign{\medskip}
\end{tabular}
\end{table}
The effect of blood flow direction is further investigated for the physiologically realistic case corresponding to 
$\widetilde{m}_w=10~\widetilde{m}_p$.
To better illustrate the involved heat transfer mechanisms, Fig.~\ref{fig:flux} compares the temperature contour fields and the corresponding heat flux lines for the four considered orientations of the blood velocity.
They refer to a treatment time $\mathrm{t=950~s}$, that is the last temperature peak deriving from the laser heat source administration.
It is important to remark that the heat flux lines are based on the Fourier's law for Pennes model, \emph{i.e.}, $\mathbf{q}_t=-\lambda_t \boldsymbol{\nabla}T_t$,
while $\mathbf{q}_t=-\lambda_t \boldsymbol{\nabla}T_t+\rho_b c_{p,b} \mathbf{v}_h (T-T_0)$ is considered for Wulff model, adopting $T_0=310.15~K$. 
It is worth noting that $\mathbf{q}_t$ vector representation would not be possible within the original interpretation of the Wulff formulation, since the heat flux vector would also depend on the spatial distribution of the metabolic efficiency, a quantity that is not solved for independently. Treating metabolic heat generation as a separate volumetric source therefore not only provides a physically clearer interpretation, but also yields a heat flux vector that can be directly evaluated and visualized.\\
Looking at Fig.~\ref{fig:fPennes}, it is evident that when the Pennes model is employed, the heat flux is primarily aligned with the vertical (negative) direction near the domain’s axis of symmetry.
In contrast, in the second portion of the tumor ($\mathrm{x > 3mm}$), the vector $\mathbf{q}_t$ forms an angle of approximately $45^\circ$ with the vertical axis.
Considering the definition of the heat flux vector given by Wullf, is evident that the blood velocity orientation has a strong impact on the thermal field developed into the tumor. 
In fact, when $\mathbf{v}_h$ is equal to $v_h~\mathbf{\hat{i}}$ (Fig.~\ref{fig:fUpos}), $\mathbf{q}_t$ has an additional horizontal component with respect to the reference case using Pennes equation, promoting the heat spread within the domain which, in turns, produces a rise in temperature in the zone to be treated and in the surrounding tissue.
Conversely, a blood velocity in horizontal, negative direction strongly opposes to the temperature gradient, especially in the second portion of the tumor, as noticeable in Fig.~\ref{fig:fUneg}.
This effect is even amplified by imposing $v_h~\mathbf{\hat{j}}$. 
In fact, the blood contribution in $\mathbf{q}_t$ limits the heat spread into the domain, bounding the rise in temperature in the upper part of the tumor (see Fig.~\ref{fig:fvpos}), with a consequent enhancement of the convective heat flux on the skin surface.
Obviously, $\mathbf{v}_h$ characterized by a vertical, negative direction, results in an opposite effect.
In fact, as shown in Fig.~\ref{fig:fVneg}, all of the melanoma reaches a temperature such as to be in the coagulative regime.
This is related to the fact that blood perfusion promote the heat diffusion in the peripheral zones of the tumor.\\
The heat transfer mechanisms discussed above are consistently reflected in the local thermal response. To further quantify the influence of blood flow direction, Fig.~\ref{fig:Tab_m} and Fig.~\ref{fig:Tcd_m} report the temperature time--histories at the four monitoring points introduced in Fig.~\ref{fig:scheme}. These local temperature histories provide a quantitative counterpart to the global temperature fields and heat flux patterns discussed above. For the case of $\mathbf{v}_h=v_h,\mathbf{\hat{i}}$, the Wulff model predicts higher temperatures throughout the tumor than the classical Pennes formulation.
%
Imposing a horizontal blood velocity in the negative direction results in temperature values predicted by the Wulff model that are considerably lower than those provided by the Pennes bio--heat equation, as confirmed by the plots in Fig.~\ref{fig:Tab_m}.
For what concerns the effect of a vertical blood velocity, two contrasting behaviors can be noted.
Indeed, $\mathbf{v}_h=v_h~\mathbf{\hat{j}}$  results in a significant reduction in temperature within the domain when compared to the reference scenario that utilizes the Pennes blood perfusion model.
Additionally, the predicted temperature never exceeds the coagulation regime threshold.
This trend is clearly visible in temperature time-histories showed in Fig.~\ref{fig:Ta_m}.
On the other hand, $\mathbf{v}_h=-v_h~\mathbf{\hat{j}}$ makes the contribution of the blood enthalpy flux to the total heat flux aligned with the one due to the temperature gradient.
As a consequence, the temperature of the tumor is higher than any other configuration studied in this work.
In this context is important to highlight that, for the latter configuration, even the evaporation condition is reached in point A of the tumor (see Fig.~\ref{fig:Ta_m}).\\
Overall, the combined interpretation of the spatial and temporal thermal fields demonstrates that the differences between the Pennes and Wulff formulations originate from the directional blood enthalpy transport accounted for only by the latter. While moderate discrepancies are observed when $\widetilde{m}_w=\widetilde{m}_p$, adopting a physiologically realistic Darcy velocity ($\widetilde{m}_w=10\,\widetilde{m}_p$), leads to markedly different predictions in terms of both the thermal field and treatment outcome.\\
\begin{figure}[htbp]
 \centering
 \subfigure[\label{fig:Ta1000} Point A.]
 {\includegraphics[width=0.45\textwidth, trim=2mm 2mm 2mm 2mm, clip]{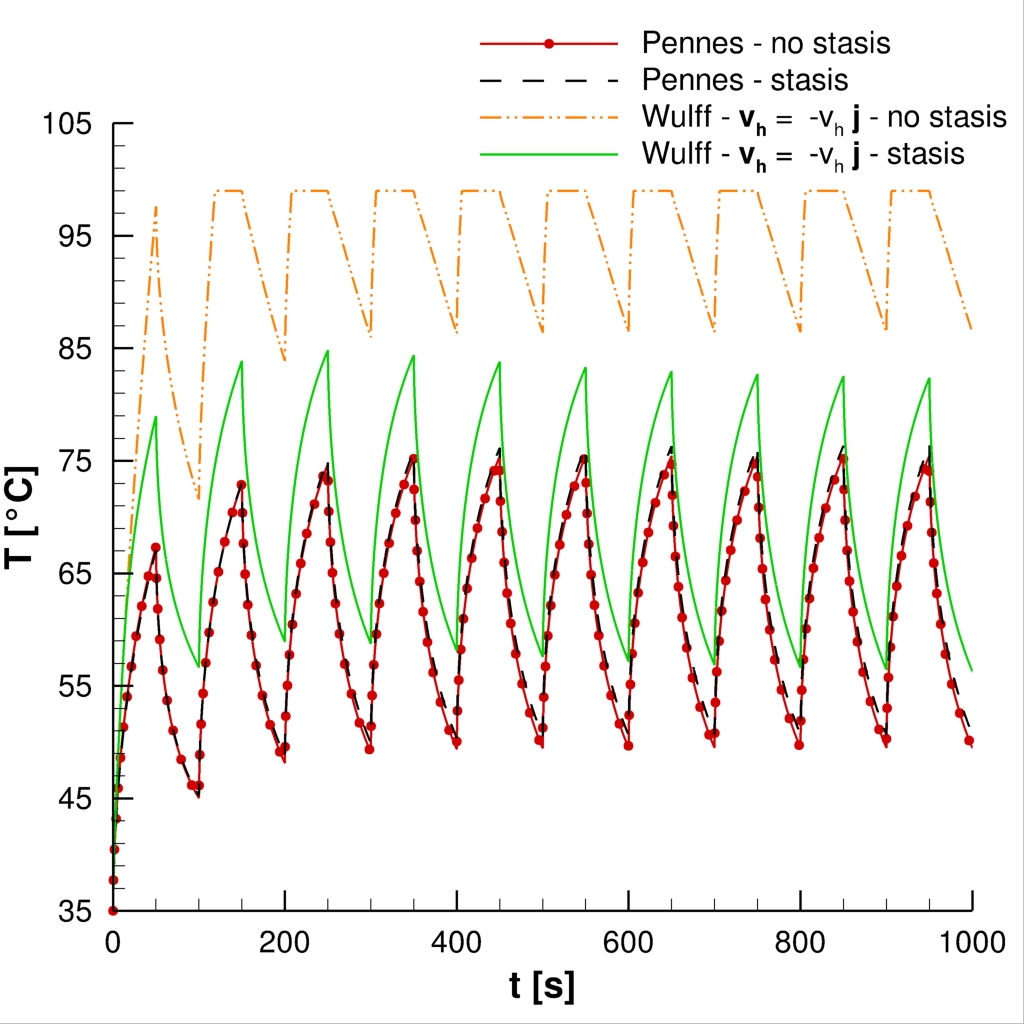}}
\hfill
  \subfigure[\label{fig:Tb1000} Point B.]
 {\includegraphics[width=0.45\textwidth, trim=2mm 2mm 2mm 2mm, clip]{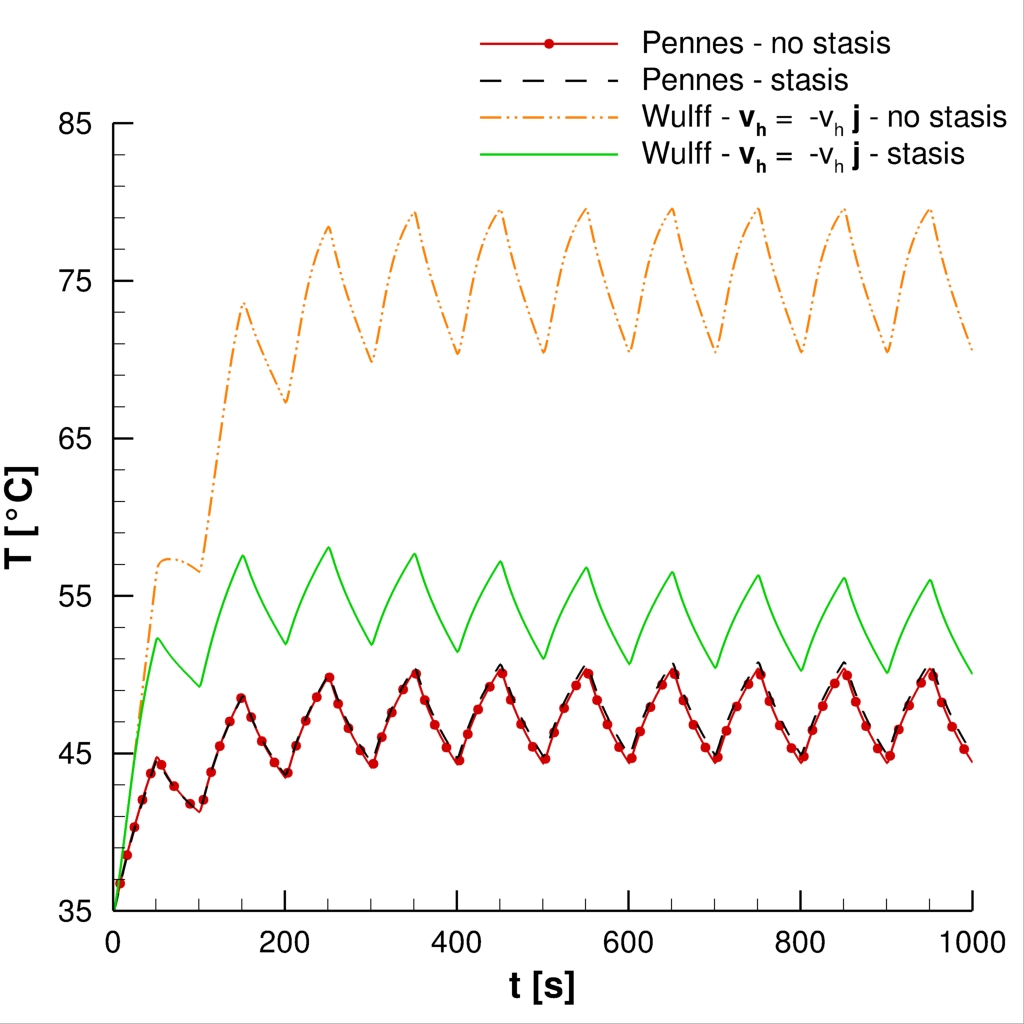}}
\caption{Time-history of temperature in point A and B, $\tilde{m}_w  = 10~\tilde{m}_p$. Venous stasis effect.}
\label{fig:Ta_m}
\end{figure}
\noindent Tab.~\ref{tab:E} reports the obtained treatment efficiency $\varepsilon_L$ for this setup. Specifically, adopting the Pennes formulation as the blood perfusion model results in a treatment efficiency of 63.8\%, as already discussed. In contrast, the Wulff model yields varying outcomes depending on the direction of the blood flow velocity. When the velocity vector $\mathbf{v}_h$ opposes the temperature gradient, the tumor region reaching the threshold $\Omega > 1$ is smaller than in the Pennes case, resulting in a reduced efficiency of 18.8\% for $\mathbf{v}_h = -v_h~\mathbf{\hat{i}}$.
Moreover, when the blood flows vertically upward,~\emph{i.e.} $\mathbf{v}_h = v_h~\mathbf{\hat{j}}$, the treatment proves completely ineffective, with $\varepsilon_L = 0$. Conversely, when $\mathbf{v}_h$ supports heat diffusion into the tumor (\emph{i.e.} $\mathbf{v}_h = v_h~\mathbf{\hat{i}}$ and $\mathbf{v}_h = -v_h~\mathbf{\hat{j}}$), the entire tumor reaches the $\Omega > 1$ threshold, leading to a treatment efficiency of 100\%. However, it is important to note that in these last two cases, the extent of irreversible thermal damage also affects healthy tissue. As shown in Tab.~\ref{tab:E}, the percentage of healthy tissue irreversibly damaged, $U_L$, increases to 37.5\% and 63.0\% for $\mathbf{v}_h = v_h\mathbf{\hat{i}}$ and $\mathbf{v}_h = -v_h~\mathbf{\hat{j}}$, respectively. Overall, both the efficiency and the side effects of the treatment are highly sensitive to the choice of model and the direction of perfusion velocity. The Pennes model offers moderate efficiency (approximately 64\%) without collateral damage. In contrast, the Wulff model yields full efficiency when the velocity is along the positive \textit{x}-axis but introduces moderate side effects (about 36\%). Reversing the flow along the same axis lowers efficiency to 19\% with no side effects. Along the \textit{y}-axis, upward flow results in no therapeutic effect, while downward flow restores full efficiency but at the cost of substantial damage to healthy tissue (approximately 63\%). These results highlight the crucial role of blood flow direction in determining both the efficacy and safety of the treatment.

\begin{table}
\caption{Treatment efficiency and undesired effect, $\widetilde{m}_w=10~\widetilde{m}_p$}
\label{tab:E}
\centering
\begin{tabular}{ccccccc}
\hline\noalign{\medskip}
& Pennes & Wulff & Wulff & Wulff & Wulff\\
&& $\mathbf{v_h}=v_h~\mathbf{\hat{i}}$ & 
$\mathbf{v_h}=-v_h~\mathbf{\hat{i}}$ &
$\mathbf{v_h}=v_h~\mathbf{\hat{j}}$ & 
$\mathbf{v_h}=-v_h~\mathbf{\hat{j}}$
\\\hline\noalign{\medskip}
$\mathrm{\varepsilon_L}$~~[\%] & 63.8 & 100.0 & 18.8 & 0.0 & 100.0\\
$\mathrm{U_{L}}$~[\%]    &   0.0 &  35.7 &  0.0 & 0.0  &  63.0\\
\end{tabular}
\end{table}

The results discussed so far have been obtained without accounting for the effects of vascular stasis. However, vascular stasis represents an important phenomenon during thermal therapies, since the progressive destruction of the microvasculature reduces blood circulation and, consequently, blood heat transfer. While this effect has been extensively incorporated into Pennes based bio-heat models through perfusion attenuation formulations, at the time of this writing no analogous extension has been proposed for the Wulff model, despite its explicit dependence on blood convection. To address this issue, the modified Wulff formulation introduced in Sec.~\ref{sec:gov} is finally assessed. Fig.~\ref{fig:Ta_m} compares the Wulff predictions obtained with and without accounting for venous stasis. As expected, the progressive reduction of blood velocity weakens the convective transport of thermal energy and, consequently, limits the temperature rise within the tumor. The effect is particularly evident after the first heating cycles, when thermal damage becomes significant. Although the modified formulation still predicts higher temperatures than the Pennes model, the discrepancy is considerably reduced, indicating that vascular stasis partially mitigates the directional effects introduced by the Wulff formulation. More importantly, the maximum temperature no longer reaches the tissue evaporation threshold. This result suggests that accounting for vascular stasis not only reduces the predicted temperatures but may also prevent the onset of non-physical overheating in configurations where the directional blood flow contribution is particularly strong.
\subsection{Role of blood flow directionality}
It is worth noting that the dimensionless coefficient governing the convective
transport term appearing in the Wullf equation's dimensionless form can be conveniently rewritten as
\begin{equation}
Pe_b=\widetilde{m}_wL
=
\frac{\rho_b c_{p,b}|\mathbf v_h|L}
{\lambda_t}~,
\label{eq:Peb}
\end{equation}
which represents the ratio between blood perfusion induced heat transport and
thermal diffusion. The parameter defined in eq.~\ref{eq:Peb} is hereafter referred
to as the \emph{biological Peclet number}. Consequently, eq.~\ref{eq:wulff_adim}
can be rewritten as

\begin{equation}
\frac{1}{Fo}\frac{\partial \widehat{T}_t}{\partial \widehat{t}}
=
\widehat{\nabla}^2\widehat{T}_t
-
Pe_b^2
\,
\widehat{\mathbf v}_h\cdot  \widehat{\NABLA}\widehat{T}_t
+
\widehat{G}_m~.
\label{eq:WulffNDPe}
\end{equation}
The biological Peclet number measures the magnitude of directional blood enthalpy transport relative to thermal conduction within the tissue. For $Pe_b \ll 1$, thermal diffusion dominates the transport process, while the
directional transport of thermal energy associated with blood perfusion becomes
progressively less important. Under these conditions, the influence of blood flow directionality on the temperature field is expected to be limited.
Conversely, when $Pe_b=\mathcal{O}(1)$ or larger, blood perfusion induced
enthalpy transport becomes comparable to thermal diffusion. Under these
conditions, the orientation of the blood flow may significantly alter the
temperature field and, consequently, the predicted extent of thermal damage.\\
This interpretation is fully consistent with the numerical results presented in this work. For the first set of simulations, in which $\widetilde{m}_w=\widetilde{m}_p=91.562$, the corresponding biological Peclet number is $Pe_b \simeq 0.45$. Therefore, conductive heat transfer remains the dominant transport mechanism and only moderate differences between the Pennes and Wulff predictions are observed. In contrast, adopting the physiologically realistic Darcy velocity 
($|\mathbf{v}_h|=10^{-4}~\mathrm{m/s}$) yields $Pe_b \simeq 4.5$. In this regime, directional blood perfusion induced heat transfer becomes comparable to thermal diffusion, explaining the pronounced dependence of the Wulff solution on the orientation of the blood flow and the substantial discrepancies with respect to the Pennes formulation. Indeed, depending on the flow orientation, the predicted thermal response ranges from an ineffective treatment to complete tumor coagulation, with local tissue temperatures even reaching the evaporation threshold.\\
The numerical results suggest that the biological Peclet number is more than a convenient dimensionless parameter. It also provides a simple physical interpretation of the differences observed between the Pennes and Wulff formulations.
When $Pe_b \ll 1$, heat transfer is mainly governed by thermal diffusion and the
directional transport associated with blood perfusion plays only a secondary role.
%
As $Pe_b$ increases, the contribution of blood borne heat transfer becomes
progressively more important. In this regime, the orientation of the Darcy velocity
starts to influence the thermal field and the differences between the two formulations
become increasingly evident. 
The biological Peclet number therefore defines the asymptotic limit in which the Wulff equation degenerates into a diffusion dominated formulation. In this limit the directional transport term becomes a higher--order perturbation so that neglecting blood flow directionality is expected to introduce only minor errors. Consequently, isotropic perfusion models such as the Pennes equation are able to provide an adequate approximation of the directional blood flow effects.
In other words, the biological Peclet number should not be interpreted solely as a measure of the relative importance of blood perfusion with respect to thermal diffusion. Indeed, it also provides a measure of the role of directional blood enthalpy transport.\\
While additional experimental validation will be required to further assess the predictive capabilities of directional bioheat models, the biological Peclet number already provides a useful physical guideline for selecting the appropriate formulation. When $Pe_b$ is sufficiently small, the Pennes formulation is expected
to provide predictions close to those of the Wulff model. On the other hand, for
$Pe_b=\mathcal{O}(1)$ or larger, directional blood transport can no longer be neglected,
making the Wulff formulation a more appropriate modelling framework.
%
%
%

\section{Conclusion} \label{sec:concl}
This work has revisited the Wulff bio–heat transfer model and assessed its implications for laser-induced thermal therapy through a systematic comparison with the classical Pennes formulation. From the theoretical point of view, the physical assumptions underlying the Wulff equation have been clarified, showing that the same governing equation can be recovered by treating metabolic heat generation as an independent volumetric source rather than relating it to spatial variations of metabolic efficiency. This alternative interpretation preserves the mathematical structure of the original formulation while avoiding the associated conceptual inconsistency.\\
The numerical simulations demonstrate that accounting for blood flow directionality may substantially modify both the predicted temperature field and the extent of thermal damage. The proposed biological Peclet number provides a simple physical interpretation of these results by quantifying the relative importance of directional blood enthalpy transport with respect to thermal diffusion. When the biological Peclet number is small, neglecting blood flow directionality introduces only minor differences and isotropic bio--heat formulations are expected to provide an adequate approximation. As the biological Peclet number approaches or exceeds unity, however, directional blood transport becomes an increasingly important heat transfer mechanism, making the predicted thermal response strongly dependent on the orientation of the blood flow.
The present study does not demonstrate that the Wulff formulation is intrinsically more accurate than the Pennes model, since suitable experimental evidence for such a comparison is still unavailable. Nevertheless, the results indicate that the directional component of blood heat transfer cannot always be neglected and that its influence should be assessed on the basis of the underlying transport regime rather than assumed a priori.   \\
Finally, the proposed extension accounting for venous stasis progressively attenuates the directional convective transport as thermal damage develops, reducing the excessive temperature rise predicted by the original Wulff formulation. More broadly, the present work provides a physically consistent framework for interpreting directional bio--heat models and may serve as a basis for future experimental validation and for the development of more realistic formulations accounting for the complex nature of blood heat transport.
\bibliographystyle{elsarticle-num}
\bibliography{main}

\end{document}